\documentclass[11pt,a4paper]{article}
\pdfoutput=1 
\usepackage{jheppub}
\usepackage{latexsym,amsfonts,amsmath,amssymb}
\usepackage{color}
\usepackage{caption}
\usepackage{subcaption}
\usepackage{mathtools}
\usepackage[normalem]{ulem}
\usepackage{url}
\usepackage{slashed}
\def\ve{\varepsilon}
\def\vp{\varphi}

\newcommand{\CC}{\mathcal{C}}

\newcommand{\CF}{\mathcal{F}}

\newcommand{\CK}{\mathcal{K}}

\newcommand{\CN}{\mathcal{N}}

\newcommand{\CV}{\mathcal{V}}
\newcommand{\CQ}{\mathcal{Q}}

\newcommand{\IZ}{\mathbb{Z}}

\newcommand{\ndt}{\noindent}

\def\i{\mathrm{i}}

\def\p{\partial}

\def\bea{\begin{eqnarray}}
\def\eea{\end{eqnarray}}
\def\be{\begin{equation}}
\def\ee{\end{equation}}
\def\bse{\begin{subequations}}
\def\ese{\end{subequations}}

\newcommand{\bem}{\begin{pmatrix}}
\newcommand{\eem}{\end{pmatrix}}

\renewcommand{\=}{\;  = \;}
\def\+{\, + \,}

\def\wt{\widetilde}
\def\wh{\widehat}

\def\rt2{\sqrt{2}}

\def\s{\sigma}
\def\g{\gamma}

\def\a{\alpha}

\def\m{\mu}
\def\n{\nu}

\def\ve{\varepsilon}

\renewcommand{\v}{\varphi}

\newcommand{\nv}{n_\text{v}}
\newcommand{\nh}{n_\text{h}}
\newcommand{\NV}{N_\text{V}}
\newcommand{\eD}{\mathcal{D}}
\newcommand{\eS}{\mathcal{S}}
\newcommand{\ic}{\mathit{c}}
\newcommand{\q}{\vec{q}}

\newcommand{\eL}{L}
\newcommand{\defeq}{\; \coloneqq \;} 

\title{Quantum entropy of BMPV black holes and \\ the topological M-theory conjecture}

\author{Rajesh Kumar Gupta$^a$,}
\author{Sameer Murthy$^b$,}
\author{Manya Sahni$^b$}

\affiliation{$a$ Department of Physics, Indian Institute of Technology Ropar,
Rupnagar, Punjab 140001, India}

\affiliation{$b$ Department of Mathematics, King's College London,  The Strand, London WC2R 2LS, U.K}

\emailAdd{rajesh.gupta@iitrpr.ac.in}
\emailAdd{sameer.murthy@kcl.ac.uk}
\emailAdd{manya.sahni@kcl.ac.uk}

\abstract{
We present a formula for the quantum entropy of supersymmetric five-dimensional 
spinning black holes in M-theory compactified on~$CY_3$, i.e.,~BMPV black holes. 
We use supersymmetric localization in the framework of off-shell five dimensional~$N=2$ 
supergravity coupled to~$I = 1\,,\,\dots\,,\,N_V + 1$ off-shell vector multiplets. 
The theory is governed at two-derivative level by the symmetric tensor~$\CC_{IJK}$ 
(the intersection numbers of the Calabi-Yau) and at four-derivative level by the 
gauge-gravitational Chern-Simons coupling~$\ic_I$ (the second Chern class of the Calabi-Yau).
The quantum entropy is an~$N_V + 2$-dimensional integral parameterised by one 
real parameter~$\v^I$ for each vector multiplet and an additional parameter~$\v^0$ 
for the gravity multiplet. The integrand consists of an action governed completely 
by~$\CC_{IJK}$ and~$\ic_{I}$, and a one-loop determinant. Consistency with the 
on-shell logarithmic corrections to the entropy, the symmetries of the very special 
geometry of the moduli space, and an assumption of analyticity constrains the 
one-loop determinant up to a scale-independent function~$f(\v^0)$.
For~$f=1$ our result agrees completely with the topological 
M-theory conjecture of Dijkgraaf, Gukov, Nietzke, and Vafa for static black holes at 
two derivative level, and provides a natural extension to higher derivative corrections. 
For rotating BMPV black holes, our result differs from the DGNV conjecture at the 
level of the first quantum corrections.
}

\begin{document}

\maketitle

\section{Introduction \label{sec:introduction}}

In the last two decades we have seen a great deal of progress on understanding and calculating 
quantum effects on the entropy of supersymmetric black holes (BH) (see~\cite{Mandal:2010cj} for a review). 
In particular, the work of~\cite{LopesCardoso:1998tkj} on higher-derivative corrections to the 
entropy of supersymmetric black holes in supergravity, and the subsequent OSV 
conjecture~\cite{Ooguri:2004zv} inspired a great deal of work which has led to detailed insights into the 
nature of BH microstates in string theory. 
The remarkable OSV conjecture relates the entropy of supersymmetric black holes in a 
four-dimensional compactification of Type II string theory on a Calabi-Yau three-fold 
to the free energy of topological strings on the same CY$_3$. 
One particularly fruitful way to interpret the conjecture began from the work of Sen~\cite{Sen:2008vm}, which 
gave a precise notion of quantum BH entropy as a functional integral in the near-horizon 
geometry in the context of the AdS$_2$/CFT$_1$ correspondence. 
This line of investigation eventually led to the development of exact formulas for the 
gravitational quantum entropy for BHs in string theory, and a derivation of the OSV formula 
(in its most precise form given in~\cite{Denef:2007vg}) 
from the macroscopic point of view using localization in four-dimensional 
supergravity~\cite{Dabholkar:2010uh}.\footnote{In string compactifications with extended supersymmetry, 
one can go further to understand the non-perturbative effects as 
well~\cite{Dabholkar:2011ec,Dabholkar:2014ema,Chowdhury:2019mnb}. 
}

\vskip 0.2cm

It is somewhat less well-known that there is an equally remarkable conjecture of 
Dijkgraaf-Gukov-Nietzke-Vafa (DGNV)~\cite{Dijkgraaf:2004te} 
for the quantum entropy of supersymmetric black holes 
in \emph{five-dimensional} asymptotically flat space. 
The idea of~\cite{Dijkgraaf:2004te} is that the fundamental fields are differential forms
of various degrees and (at least some aspects of) the metric and spacetime arise from 
these forms. The fundamental principle is taken to be the Hitchin functional or the 
form-theories of gravity~\cite{Hitchin:2000jd,Hitchin:2001rw}, and the paper~\cite{Dijkgraaf:2004te} 
conjectures the existence of a theory called topological M-theory which unifies these 
different form theories. The associated spacetime is 7-dimensional spacetime,
and the theory reduces to the topological A- and B-models on a Calabi-Yau three-fold 
in appropriate limits. 

One motivation in~\cite{Dijkgraaf:2004te} for the emergence of the metric from 
cohomology (or charges) is the BH attractor mechanism~\cite{Ferrara:1996dd,Ferrara:1996um}, 
and one important
consequence (Section 8) is conjectures for quantum generalizations of the BH attractor. 
The classical limit of the B-model (or Type IIB) version of the conjecture, 
which is formulated in terms of the holomorphic Hitchin functional, yields the OSV formula 
relating the quantum 4d black hole entropy to the topological string partition function. 

In this paper we focus on the A-model version of the conjecture which relates to 
M-theory compactified on the CY$_3$. This version is based on the symplectic Hitchin 
functional, and proposes a quantum generalization of the attractor mechanism
for supersymmetric black holes in five-dimensional asymptotically flat space, 
i.e.~BMPV black holes~\cite{Breckenridge:1996is}. 
The aim of this paper is to obtain a derivation of this aspect of 
the topological M-theory conjecture from the gravitational point of view.
In particular, we present a formula for the exact quantum entropy of five-dimensional black holes 
using localization in off-shell five-dimensional supergravity. 
This formula agrees precisely with the formula of~\cite{Dijkgraaf:2004te} 
for non-rotating black holes. For rotating black holes, the two formulas agree at leading order, 
and our results suggest  
a modification of~\cite{Dijkgraaf:2004te} in order for it to be consistent with macroscopic results 
beyond the leading order.
In the rest of this introductory  section, we  
summarize the idea of our calculation, the extent to which we complete the 
derivation and the remaining gaps.

\vskip 0.4cm

It is useful to first recall the form of the (perturbatively) exact
quantum entropy formula for black holes in four-dimensional supergravity
found using localization methods~\cite{Dabholkar:2010uh}. We have 
\be \label{4dQE} 
Z^\text{4d}(q,p) \= \int\prod_{I = 0}^{\nv+1} \,d \v^I\,
\exp \Bigl(- \pi q_I \v^I +  4 \pi \, \text{Im}  F\Bigl(\frac{\v^I+ \i \, p^I}{2} \Bigr)  \Bigr)\,Z^\text{4d}_{\text{1-loop}}(\v^I)\,,
\ee
where~$F$ is the holomorphic prepotential of~$\CN=2$ four-dimensional supergravity. 
Here~$\v^I$ are the coordinates of the localization manifold~\cite{Dabholkar:2010uh,Gupta:2019xac}, 
the imaginary part of the prepotential is the value of the supergravity action on the 
localization manifold~\cite{Murthy:2013xpa},
the Legendre transform arises because of the boundary conditions on AdS$_2$ which necessitate  
the presence of a Wilson line at the boundary~\cite{Sen:2008vm}, and the 
one-loop determinant is due to the fluctuations in the non-BPS directions around the localization manifold.
This determinant was calculated in~\cite{Murthy:2015yfa, Gupta:2015gga,Jeon:2018kec}, 
and takes the following form, 
\be \label{1loopfinal}
Z^\text{4d}_\text{1-loop} \= \exp\Bigl(-\CK(\phi^{I}) \bigl(2 -\frac{\chi}{24} \bigr) \Bigr) \, , \qquad \chi \= 2(\nv+1-\nh) \,,
\ee
where~$\nv$ and~$\nh$ are the number of vector and hyper multiplets, respectively. 
The formula~\eqref{4dQE} agrees precisely with the OSV formula as interpreted by 
Denef-Moore~\cite{Denef:2007vg} (where~$\chi$ is interpreted as 
the Euler characteristic of the~CY$_3$, and the 
supersymetric black hole is a bound state of branes wrapping 
supersymmetric cycles in the $CY_3$-fold.
Now we want to follow the same route to calculate the exact quantum entropy for 
five-dimensional black holes.

\vskip 0.4cm

Before we present our results, we should immediately note that the paper~\cite{Gomes:2013cca}
presented a formula for the quantum entropy of 5d black holes, making heavy use of 
off-shell 4d/5d connection~\cite{Banerjee:2011ts}. 
The formula was presented in terms of the four-dimensional prepotential, and 
was obtained as follows. The four-dimensional off-shell BPS solutions 
found in~\cite{Dabholkar:2010uh} was first lifted to five dimensions, 
and then the two-derivative off-shell action on these solutions was calculated.
The classical action is essentially equal to the four-dimensional action, as 
consistent with the 4d-5d connection, and therefore one obtains a very similar looking 
formula as~\eqref{4dQE}.  
However,~\cite{Gomes:2013cca} left open the issue of higher-derivative corrections in the off-shell theory 
which is generically present (these terms vanishes for the special case of torus compactification). 
There is a also subtle issue of boundary terms in the action which was left open.   
Further, the quantum corrections to the leading-order entropy in the localization 
formalism was not treated. 
In fact, quantum effects are also generically present in~$\CN=2$ theories---and also in the maximally 
supersymmetric theory for a certain scaling of the angular momentum. Without these effects properly 
taken into account one does not have a correct quantum formula beyond the semiclassical regime. 
In the present paper we take into account the four-derivative supersymmetric
invariants in 5d off-shell supergravity~\cite{Banerjee:2011ts, Hanaki:2006pj}
including a careful analysis of boundary terms, and begin a discussion of the one-loop quantum corrections.

\vskip 0.4cm

In this paper we take a manifestly five-dimensional approach to this problem. Consider 
five-dimensional~$N=2$ supergravity coupled to~$N_{V}+1$ vector multiplets. 
In five-dimensional supergravity we do not have a holomorphic prepotential but, instead, 
very special geometry in the vector multiplet sector. At the two-derivative
level this is governed by a real function~$C(\sigma)=C_{IJK} \sigma^I \sigma^J \sigma^K$, 
where~$\sigma^I$ are the real scalars in the vector multiplets and~$C_{IJK}$ is the completely 
symmetric three-form which arises as the triple intersection number of four-cycles 
in the CY$_3$ compactification. 
At higher-derivative level, the known perturbative corrections are governed by the second Chern class~$c_{2I}$.
Our (perturbatively) exact quantum entropy formula is presented in 
Section~\ref{sec:quantent}. The quantum entropy of a supersymmetric black hole solution 
in this theory carrying (electric) charges~$q_I$ and angular momentum~$J$ is  
\be \label{Eqn:Zqu1}
\begin{split}
\exp \bigl(S^\text{qu}(J,q^I) \bigr) & \= \int\prod_{I = 0}^{N_{V}+1} \,d\varphi^I\,
\exp \bigl(S_{\text{Ren}}(\varphi^0, \varphi^I; J, q^I) \bigr) \,Z_{\text{1-loop}}(\varphi^0, \varphi^I; J, q^I) \,,\\
S_{\text{Ren}}(\varphi^0, \varphi^I; J, q^I) & \=  \pi q_{I}\varphi^{I} + \pi J \varphi^{0} 
-2\pi  \frac{\CC(\varphi)}{1+(\varphi^{0})^{2}}  + \frac{\pi}{2} \frac{\ic_{I}\varphi^{I}}{1+(\varphi^{0})^{2}} \,.
\end{split}
\ee
The various pieces in this formula mirror the four-dimensional derivation closely. The real parameters~$\v^I$
label the localization manifold, the exponent inside the integral is the classical action of the theory on this
manifold. The parameter~$\v^0$ arises in the gravity multiplet sector as can be seen by the fact that it 
couples to the angular momentum. 
The one-loop determinant arises from the off-shell non-BPS fluctuations. 
The one-loop determinant can be constrained by comparing the expansion of~\eqref{Eqn:Zqu1} to the 
on-shell calculations of leading logarithmic corrections to the 
Bekenstein-Hawking entropy~\cite{Sen:2011ba, Sen:2012cj}. 
As we discuss in Section~\ref{sec:one-loop} this constraint reduces the one-loop determinant 
to be a function of~$\v^0$ alone. 

The topological M-theory conjecture of~\cite{Dijkgraaf:2004te} contains a very similar formula 
as~\eqref{Eqn:Zqu1}. The discussion of~\cite{Dijkgraaf:2004te} applies only to the two-derivative
supergravity, and at this level Equation~\eqref{Eqn:Zqu1} agrees with the entropy 
formula given in that paper for non-rotating black holes. 
The formula~\eqref{Eqn:Zqu1} suggests a way to extend the ideas of~\cite{Dijkgraaf:2004te} 
to higher-derivative level.
When~$J\neq0$ our formula differs from the conjecture of~\cite{Dijkgraaf:2004te} 
(still keeping two-derivatives but at the non-linear level).
One aesthetically pleasing point about~\eqref{Eqn:Zqu1} is the manner in which gravity 
avoids the introduction of a non-analytic square-root. 
The square-root non-linearity of the classical entropy is, instead, introduced by the introduction 
of~$\v_0$ in denominator of the formula, which does not cause any singularity at the origin. 
We discuss the details of the comparison in Section~\ref{sec:topMthy}.

\vskip 0.4cm

A separate motivation for our investigations comes directly from the macroscopic supergravity analysis. 
The black hole solutions mentioned above are static solutions of four-dimensional supergravity. 
One can try to extend this understanding to higher-dimensional black holes and, 
in particular, to spinning black holes which come with a whole new set of subtleties. 
The semiclassical entropy of these black holes including higher-derivative corrections 
have been studied in~\cite{Castro:2007ci,Castro:2008ne,deWit:2009de}.
In particular, the authors of~\cite{Castro:2007ci} reduce the five-dimensional 
Chern-Simons term to four dimensions using the 4d/5d connection. 
They find that the Chern-Simons terms reduce to gauge invariant terms plus a total derivative. 
After removing this total derivative term from the from the 5d action,
the authors find agreement between the semi-classical entropy of the 4d and the 5d black holes.

Here we check this statement  directly in the off-shell  two-derivative Lagrangian without 
reference to the BH solution, i.e.,~there are no other sources of discrepancy between the 
off-shell 4d and 5d Lagrangians except for the total derivative term coming from the actual CS term.
At the four-derivative level, however, while we have been able to isolate the total derivative terms 
which account for the difference between the 4d and the 5d on-shell actions, some subtleties regarding 
total derivative terms for the auxiliary fields remain in the off-shell theory, as we 
discuss in Sections~\ref{sec:TDterms} and~\ref{sec:nonF}.

\vskip 0.4cm

Finally we note that there are many interesting directions to follow. From the point of view of 
the form theories of gravity, it would be interesting to understand the inclusion of 
higher-derivative terms proportional to the second Chern class of the~$CY_3$-fold. 
On the macroscopic side it would be nice to calculate the non-linear form of the 
one-loop determinant even for small charges. One can then compare our results 
to known microscopic formulas for five-dimensional black 
holes including higher order corrections~\cite{Castro:2008ys,Dabholkar:2010rm, Dabholkar:2012nd}. 
If we understand non-perturbative effects in the quantum entropy one can 
eventually hope to make contact with Gopakumar-Vafa-type counting 
formulas~\cite{Gopakumar:1998ii,Gopakumar:1998jq}.
In another direction, there has been recent progress on understanding the gravitational entropy 
of supersymmetric asymptotically AdS black holes from the boundary of AdS space 
in the context of AdS/CFT 
correspondence~\cite{Cabo-Bizet:2018ehj,Cassani:2019mms,Bobev:2019zmz,BenettiGenolini:2019jdz,
Kantor:2019lfo,David:2020ems,Bobev:2020egg,Bobev:2020zov}. 
It would be interesting to extend these calculations to a full gravitational path integral,
this would involve developing off-shell localization techniques to gauged 
supergravity (see~\cite{Dabholkar:2014wpa,Nian:2017hac,Hristov:2018lod,Hristov:2019xku} 
for some results in this direction). 

\vskip 0.4cm

The plan of the paper is as follows. In Section~\ref{sec:action} we discuss the five-dimensional 
off-shell action and make a detailed comparison with the reduced action in four-dimensions including 
total derivative terms. In Section~\ref{sec:Sclass} we review the quantum entropy formalism 
and some puzzles regarding the definition of charge entering the formalism. In Section~\ref{sec:locaction}
we discuss the localization solutions and off-shell action on the localization manifold.
In Section~\ref{sec:quantent} we put all the elements together and present our formula for the 
quantum entropy of BMPV black holes and compare our results with those coming from
the topological M-theory conjecture. 
In three appendices we present some details of the five-dimensional off-shell theory, and 
evaluate the quantum entropy for a toy model.

\section{The 5d off-shell supergravity action and total derivative terms \label{sec:action}}

In this section we begin in~\S\ref{sec:4d5dreview} by reviewing the off-shell supergravity 
formalism in five dimensions, following the treatment 
of~\cite{Banerjee:2011ts}.
The action of this theory contains 
two-derivative and four-derivative supersymmetry invariants.  In~\S\ref{sec:4d5dfields}  
we summarize the \emph{off-shell 4d/5d connection}~\cite{Banerjee:2011ts} which is a map between 
four-dimensional and five-dimensional off-shell  supergravity in the context of a Kaluza-Klein 
reduction on a circle. The map relates four-dimensional
and five-dimensional field configurations so as to preserve the off-shell supersymmetry
transformations. In particular, this ensures that any four-dimensional solution of the off-shell
BPS equations lifts to a corresponding solution in five dimensions. 
Further, the map relates the off-shell actions in four and five dimensions. The relation is 
essentially that the two actions are equal on configurations related through the lift, but there 
are subtleties coming from total derivative terms which are needed to ensure 
gauge invariance in the presence of a Chern-Simons coupling in the Lagrangian.
These total derivative terms turn out to be essential to understand the relation between the 
four-dimensional and the corresponding five-dimensional BHs.
In~\S\ref{sec:TDterms} we work out the explicit details of these total-derivative terms, 
thus filling a gap in the literature. 
Finally, in \S\ref{sec:nonF} we make some comments about the relevance of 
terms that do not arise as chiral superspace integrals of a holomorphic prepotential.

\subsection{Review of off-shell action of 5d supergravity \label{sec:4d5dreview}}

We work with 5d off-shell supergravity coupled to $N_{V} + 1$ vector multiplets in the superconformal formalism. 
The gravity sector is encoded in the Weyl multiplet, whose independent fields are
\be\label{Eqn:WeylFields}
\mathbf{W} \= (e_{M}^{A}, \psi_{M}^{i}, b_{M}, V_{M i}{}^{j}; T_{AB}, \chi^{i}, D) \,.
\ee
Here the indices $A,B,\dots$, $M,N,\dots$, and $i,j,\dots$ are five dimensional flat space, 
curved space, and $SU(2)_{R}$ fundamental indices, respectively. 
The fields are the f${\ddot{\text u}}$nfbein $e_{M}^{A}$, the gravitini~$\psi_{M}^{i}$, the dilatation 
gauge field $b_{M}$, the $SU(2)_{R}$ gauge fields $V_{M i}{}^{j}$, as well as the auxiliary 
fields which are the anti-symmetric tensor~$T_{AB}$, the spinor~$\chi^{i}$, and the scalar $D$.
The field content of each five-dimensional vector multiplet is 
\be
\mathbf{V}^{I} \= (\sigma^{I}, \Omega^{i}{}^{I}, W^{I}_{M}, Y^{I}_{ij}) \,,
\label{Eqn:VecM}
\ee
where $W^{I}_{M}$ is a gauge field, $\sigma^{I}$ is a real scalar, $\Omega^{i}{}^{I}$ are the 
gaugini doublet under $SU(2)_{R}$,  
and $Y^{I}_{ij}$ are the~$SU(2)_R$ triplet auxiliary fields.
The index $I\=1, \dots ,N_{V}+1$ label the vector multiplets. 
One of the vector multiplets is a compensator multiplet used in the gauge-fixing procedure 
to obtain the on-shell theory. The on-shell theory has~$N_V$ vector multiplets and one graviphoton
vector field in the gravity multiplet.

The total 5d Lagrangian in~\cite{Banerjee:2011ts} is written as a sum of a two-derivative piece 
and a four-derivative piece
\be \label{L5d}
\eL_{5d}  \= \eL_{VVV} + \eL_{VWW} \,.
\ee
The notation, which is the same as in~\cite{Banerjee:2011ts}, stands for a Lagrangian which 
describes the supersymmeterization of the coupling of three vector multiplet fields and one 
vector multiplet with two Weyl multiplet fields, respectively. The first piece contains the 
canonical two-derivative terms and their supersymmetric completion. The second piece 
contains the four-derivative Chern-Simons term and its supersymmetric completion.
The two pieces are separately invariant under off-shell supersymmetry. 
Here and in the following presentation we display the Lagrangian including the 
factor~$E=\det (e_{M}^{A})$, so that the action is the 
integral of the Lagrangian over the spacetime manifold (up to boundary terms that we discuss in
detail below). 

The two-derivative bosonic Lagrangian of the Weyl multiplet coupled to~$\NV$ vector 
multiplets is
\be\label{Eqn:5dLVVV}
8\pi^{2} \eL_{VVV} \= \eL_{1} + \eL_{2} + \eL_{3}\,,
\ee
with
\be \label{Eqn:L1} 
\begin{split}
{\eL}_{1} &=  3E \, \mathcal{C}_{IJK} \, \sigma^{I}
\Bigl(\frac{1}{2}\mathcal D_{M}\sigma^{J}\mathcal D^{M}\sigma^{K}
+\frac{1}{4}F_{MN}{}^{J}F^{MN K}-Y_{ij}{}^{J}Y^{ij K}-3\sigma^{J}F_{MN}{}^{K}T^{MN}\Bigr)\,, \\
{\eL}_{2} &=  -\mathcal{C}(\sigma) \, E \,\Bigl( -\frac{1}{8}R-4D-\frac{39}{2}T^{2}\Bigr)\,,  \\
{\eL}_{3} &=  
-\frac{i}{8} \, \mathcal{C}_{IJK} \, \varepsilon^{MNPQR} \, W^{I}_{M}F_{NP}{}^{J}F_{QR}{}^{K}\,.
\end{split}
\ee
Here $R$ is the 5d Ricci scalar, $\mathcal{C}_{IJK}$ is the symmetric tensor of~$\CN=2$ supergravity, and 
\be
\CC(\sigma)\=\mathcal{C}_{IJK} \, \sigma^{I}\sigma^{J}\sigma^{K} \,.
\ee
In the context of M-theory compactified on a~$CY_3$-fold, and $\mathcal{C}_{IJK}$ 
is the triple intersection number of the 4-cycles in the~$CY_3$. 
A common choice REFvanProeyenTExt of gauge-fixing is~$\CC(\s)=1$.
Here it is convenient to use a slightly different choice as in~\cite{Dabholkar:2010uh} in 
which~$\sqrt{g^\text{4d}}\, \CC(\s) = 1$ where~$g^\text{4d}$ is the metric in four dimensions.
The four-derivative bosonic Lagrangian~\cite{Banerjee:2011ts} is
\be
\begin{split} \label{Eqn:5dLVWW}
& 8\pi^2\eL_{VWW} \\
&\quad  \= \frac{1}{4} \, E \, \ic_{I} \, Y^{I}_{ij} \, T^{AB} \, R_{ABk}^{j}(V) \, \varepsilon^{ki}  \\
& \qquad+ E \, \ic_{I} \, \sigma^{I}\Bigl(\frac{1}{64}R_{ABCD}(M) \, R^{CDAB}(M) 
-\frac{1}{96}R_{ABj}{}^{i}(V) \, R^{AB}{}_{i}{}^{j}(V)\Bigr)\\ 
&\qquad
-\frac{\i}{128}\varepsilon^{MNPQR} \, \ic_{I} \, W_{M}^{I}
\Bigl(R_{NPAB}(M) \, R_{QR}{}^{AB}(M) +\frac{1}{3}R_{NPj}{}^{i}(V) \, R^{QR}{}_{i}{}^{j}(V)\Bigr)\\ 
&\qquad + \frac{3}{16} \, E \, \ic_{I} \, \bigl(10\,\sigma^{I} T_{AB} - F^{I}_{AB} \bigr) \,R^{CDAB}(M) \, T_{CD}\\ 
&\qquad + E \, \ic_{I} \, \sigma^{I}\Bigl( 3T^{AB} \, \eD^C \eD_A T_{BC} \\
& \qquad\qquad\qquad\qquad - \frac{3}{2}\bigl((\eD_A T_{BC})^2
-\eD_C T_{AB}\eD^A T_{CB}\bigr)  -R_{AB}(T^{AC}T^{B}{}_{C} - \frac{1}{2}\eta^{AB}T^2)\Bigr) \\ 
&\qquad 
+ E \, \ic_{I} \, \sigma^{I}\Bigl(\frac{8}{3}D^2 + 8T^2 D-\frac{33}{8}(T^{2})^{2} +\frac{81}{2}(T_{AC}T^{BC})^2\Bigr)\\ 
&\qquad -E\, \ic_{I} \, F^{I}_{AB}\Bigl(T^{AB}D + \frac{3}{8}T^{AB}T^{2} -\frac{9}{2}T^{AC}T_{CD}T^{DB}\Bigr)\\ 
&\qquad
+ \frac{3\,\i}{4} \, E \, \ic_{I} \, \varepsilon^{ABCDE}\Bigl(F^{I}_{AB}(T_{CF}\eD^{F}T_{DE} 
+ \frac{3}{2}T_{CF}\eD_{D}T_{E}{}^{F}) -3\sigma^{I} T_{AB}T_{CD}\eD^{F}T_{FE}\Bigr) \,.
\end{split}
\ee
Here~$R_{MNi}{}^{j}(V)$ is the supersymmetric curvature, which, upon setting the fermions to zero, takes 
the usual form,
\be \label{Eqn:RVexp}
R_{MNi}{}^{j}(V) \= 2 \, \eD_{[M}V_{N]i}{}^{j} - V_{[M i}{}^{k}V_{N] k}{}^{j}\,.
\ee
$R_{MN}{}^{AB}(M)$ is the supersymmetric completion of the Weyl tensor.
It can be expressed in terms of the independent fields of the Weyl multiplet after imposing the 
conventional constraints. We present some details in Appendix~\ref{App:rem4d5dfields}. 
Upon setting the fermions to zero, we have
\be
\label{Eqn:R(M)}
R_{MN}{}^{AB}(M) \= R_{MN}{}^{AB} -\frac{4}{3} \, e_{[M}{}^{[A}R_{N]}{}^{B]} +\frac{R}{6} \, e_{[M}{}^{[A}e_{N]}{}^{B]} \,.
\ee

\subsection{Review of off-shell 4d/5d lift \label{sec:4d5dfields}}

The off-shell 4d/5d connection~\cite{Banerjee:2011ts}  is a relation between 5d and 4d off-shell supergravity theories 
when the 4d theory is the Kaluza-Klein (KK) reduction of the 5d theory on a circle. 
In this context, the 4d/5d connection provides a map between the off-shell supersymmetry equations 
as well as the respective off-shell Lagrangians. 
The 5d supergravity theory was described above. We now present the 4d theory and then the relation between them.
Throughout this section we follow the treatment of~\cite{Banerjee:2011ts}.

The 4d supergravity describes the Weyl multiplet coupled to~$\nv+1$ vector multiplets, with~$\nv=\NV+1$. 
The 4d Weyl multiplet has the following independent fields: 
\be \label{Eqn:WeylFields4d}
\mathbf{W}_{\text{4d}} \= (e^{a}_{\m},\psi^{i}_{\m},b_{\m},\mathcal V^{i}_{\m j},T_{ab},\mathcal A_{\m},\chi^{i},D) \,,
\ee 
and the 4d vector multiplets consist of 
\be \label{Eqn:VecM4d}
\mathbf{X}^{\wh{I}} \= (X^{\wh{I}}\,,\lambda^{\wh{I}}_{i}\,,A^{\wh{I}}_{\m}\,, y^{ij\wh{I}})\,, 
\qquad \wh{I} \= 0, 1, \dots , \nv \,.
\ee
The relevant 4d Lagrangian under consideration is governed by the prepotential function~$F$ which is a 
holomorphic function of the superfields,
\be \label{Eqn:Prephd}
F (X^{\wh I}, \wh A) \= -\frac{1}{2}\frac{C(X)}{X^0} - \frac{1}{2048}\frac{c_{I}X^{I}}{X^0}\widehat{A} + \dots \,,
\ee
where $\mathbf{\widehat{A}}$ is a superfield containing the Weyl squared tensor, whose lowest component is
$\widehat{A} \= (T^{-})^2$.

The function~$F$ is homogeneous of degree two, with~$X^I$ having degree one and~$\wh A$ 
having degree two. The first term on the right-hand side of~\eqref{Eqn:Prephd} is the classical 
prepotential which leads to the canonical two-derivative theory.
The second term is the leading correction which contains the four-derivative F-term and its 
supersymmetric completion. The ellipses represent higher-order terms that arise from 
instanton effects in the~$CY_3$ compactification. 
The full supersymmetric Lagrangian is
\be \label{L4d}
\eL_{4d} \= \eL_{4d}^{(0)} +  \eL_{4d}^{(1)} \,,
\ee
where  
\be \label{L4dpieces}
\begin{split}
8\pi \eL_{4d}^{(0)} & \= \Bigl(-\i\, \eD_{\mu}X^{\widehat{I}}\eD^{\mu}\overline{F}_{\widehat{I}} 
+ \i\, X^{\widehat{I}}\overline{F}_{\widehat{I}}\bigl(\frac{\widehat{R}}{6} - D\bigr)
+\frac{\i}{4}F_{\widehat{I}\widehat{J}}F^{-\widehat{I}}_{\m\n}F^{-\m\n \widehat{I}} \\
&  \qquad\qquad \quad + \frac{1}{8}\overline{X}^{\widehat{I}}N_{\widehat{I}\widehat{J}}F^{-ab\widehat{I}}T^{-}_{ab}
-\frac{1}{64}\overline{X}^{\widehat{I}}N_{\widehat{I}\widehat{J}}\overline{X}^{\widehat{J}}(T^{-})^2+h.c\Bigr) 
+\frac{1}{8}N_{\widehat{I}\widehat{J}}Y^{\widehat{I}}_{ij}Y^{\widehat{J}}_{ij}\,, \\
8\pi \eL_{4d}^{(1)}  & \= -4 \, \i\, eF_{\widehat{A}\widehat{I}}T^{-cd}\Bigl(2R(M)_{cd}{}^{ab}(F^{- \widehat{I}}_{ab} 
- \frac{1}{4}\overline{X}^{\widehat{I}}T^{-}_{ab})-\varepsilon_{ki}R(\widehat{V})_{cd}^{K}{}_{j}Y^{ij\widehat{I}}\Bigr)\\
&  \qquad \qquad\quad + 16 \,\i\, e F_{\widehat{A}}\Bigl(2R(M)^{-cd}{}_{ab}R(M)^{-ab}{}_{cd}
+R(\widehat{V})^{-abk}{}_{l}R(\widehat{V})^{-l}{}_{abk}\Bigr) \,.\\
\end{split}
\ee
Here $N_{\widehat{I}\widehat{J}} = -\i F_{\widehat{I}\widehat{J}}+ \i \overline{F}_{\widehat{I}\widehat{J}}$ 
and $a,b, \dots$ and $\m,\n, \dots$ are 4-dimensional flat-space and curved-space indices, respectively. 
We note that at two-derivative level the above F-term Lagrangian~$L^{(0)}_{4d}$ is the unique 
supersymmetry invariant. At higher-derivative level 
there are other possible supersymmetric invariants in 4d (full superspace integrals~\cite{deWit:2010za}), 
and these will play a role in our discussion of Section~\ref{sec:Sclass1}.

The KK ansatz for 5d to 4d reduction on a circle is as follows, 
\be \label{Eqn:4d5dviel}
e_{M}{}^{A}\=\begin{pmatrix}e_{\m}{}^{a}\,\, &B_{\m}\,\phi^{-1}\\0\,\,&\phi^{-1}\end{pmatrix},\qquad 
e_{A}{}^{M}\=\begin{pmatrix}e_{a}{}^{\m}\,\,&-e_{a}{}^{\n}B_{\n}\\0\,\,&\phi\end{pmatrix} \,,
\ee
with~$B_\mu$ causing the non-trivial fibration of the circle of size~$\phi^{-1}$ over the 4d base space. 
We denote the 5th coordinate by $x^5=\rho$, so that~$A=(a,5)$ and~$M=(\mu,\rho)$ in the discussion below.
In the off-shell supersymmetric theory, the 5d Weyl multiplet~\eqref{Eqn:WeylFields} reduces to the 
4d Weyl multiplet~\eqref{Eqn:WeylFields4d} and 
the vector multiplet~${\bf X}^0$, with the following relations
\be \label{Eqn:4d5dphi}
\phi \= 2|X^0|\,, \qquad B_{\mu} \= A^0_{\mu} \,.
\ee
At first sight the phase of~$X^0$ is lost in the 5d theory, but it is encoded in the auxiliary 
field~$T_{ab}$ and in $V_{Mi}^{j}$. 
As we will see below, this phase plays a role in determining the precise relation between the 4d and 5d vector fields.
The map of the other fields in the off-shell Weyl multiplet is given in Appendix~\ref{App:rem4d5dfields}. 
The 5d vector multiplet~\eqref{Eqn:VecM} reduces to the 4d vector multiplet~\eqref{Eqn:VecM4d} for~$\wh{I}=I$
with the following map between the bosonic fields, 
\be \label{Eqn:4d5dVec}
\sigma^I \= -\i \, |X^0| \, t_{-}^{I}\,,\quad \quad
Y^{I}_{ij} \= -\frac{1}{2} \, \widetilde{Y}^{I}_{ij} + \frac{1}{4} \, t^{I}_{+} \, Y^{0}_{ij}\,,\quad\quad
W^{I}_{M} \= 
\begin{cases} 
A^{I}_{\mu} -\frac{1}{2} \, t^{I}_{+} \, B_{\mu} \,, \; & M= \mu\\
-\frac{1}{2} \, t^{I}_{+}  \,, \; & M= \rho 
\end{cases} \,.
\ee
where  
\be
t^{I} \, \= \, \frac{X^{I}}{X^{0}}\,,
\quad \quad 
t^{I}_{\pm} \= t^{I} \pm \overline{t}^{I}\,.
\ee
Note that in tangent space indices we have
\be \label{Eqn:4d5dWt}
W^{I}_{a} \= A^{I}_{a}\,, \quad \quad
W^{I}_{5} \= -|X^0| \, t^{I}_{+}\,.
\ee
The relations between the field strengths, with~$\wt F_{\m\n} = \p_{\m}A_{\n}-\p_{\n}A_\m$, are
\be \label{Eqn:4d5dF}
F^{I}_{MN} \= 
\begin{cases} 
\widetilde{F}^{I}_{\m\n} + W^{I}_{\rho} \, F^{0}_{\m\n} + (B_{\n}\p_{\m} - B_{\m}\p_{\n}) W^{I}_{\rho}  \,, \; 
& (M,N)= (\mu,\nu) \\
D_{\mu}W_{\rho} \,, \; & (M,N) = (\mu,\rho) 
\end{cases} \,.
\ee
It is useful to write these equations in tangent space coordinates 
\be \label{Eqn:4d5dFta}
F^{I}_{AB} \= 
\begin{cases} 
 \widetilde{F}^{I}_{ab} + W^{I}_{\rho} \, F^{0}_{ab}\,, \; & (A,B)= (a,b) \\
 D_a W^{I}_{5} \,, \; & (A,B)= (a,5) \\
\end{cases} \,.
\ee
Other relevant details of the vector multiplet reduction are given in Appendix~\ref{App:rem4d5dfields}.

\subsection{Explicit form of total derivative terms \label{sec:TDterms}} 

The 5d action~\eqref{L5d} reduced to 4d almost agrees with the 4d action~\eqref{L4d}, 
\eqref{L4dpieces}, but not exactly. 
The difference arises purely due to a total derivative term which is needed to ensure 
gauge invariance in the presence of a Chern-Simons coupling in the Lagrangian.
We will denote the total derivative term by the superscript, TD. 
In a space where the boundary is infinitely far away and the fields drop off 
rapidly such total derivative terms can be ignored, as was done in the 
discussion of~\citep{Banerjee:2011ts}.\footnote{We thank Bernard~de~Wit for 
helpful discussions on this subject.} 
However, in the calculation of black hole entropy one needs to evaluate the action on a space 
where the boundary is at a large but finite value and then take a limit to infinity~\cite{Sen:2008vm}. 
The total derivative terms, therefore, turn out to be important for the discussion of the BH entropy 
in general. In the case of a static black hole, the leading two-derivative entropy happens to not 
depend on this term. The two-derivative entropy for \emph{spinning} black holes as well as the 
higher derivative entropy, even for the static case, are sensitive to this term. 
In the rest of this section we fill this gap in the formalism and present the total derivative terms 
in the action without any reference to a particular solution.

\vspace{0.2cm}

\ndt {\bf Two-derivative Lagrangian}

\vspace{0.2cm}

First we study the two-derivative theory given by \eqref{Eqn:5dLVVV}.
The paper~\citep{Banerjee:2011ts} gives the reduction of these three terms in detail. 
The first two are gauge-invariant and they reduce to 4d terms which we write in Appendix~\ref{App:rem4d5dLag}. 
The third term contains the Chern-Simons term given by the third line of~\eqref{Eqn:L1}, which is not 
locally gauge-invariant. This can be expanded as
\be
\begin{split}
\eL_{3} & \= -\frac{\i}{8}\mathcal{C}_{IJK}\varepsilon^{\rho\mu\nu\lambda\sigma}
W^{I}_{\rho}F_{\mu\nu}{}^{J}F_{\lambda\sigma}^{K} -\frac{\i}{2}\mathcal{C}_{IJK}
\varepsilon^{\mu\nu\lambda\sigma\rho}W^{I}_{\mu}F_{\nu\lambda}{}^{J}F_{\sigma\rho}^{K}\,, \\
&\= \frac{3 \, \i \, }{16}\mathcal{C}_{IJK}\varepsilon^{\mu\nu\lambda\sigma}t^{I}_{+}
F_{\mu\nu}{}^{J}F_{\lambda\sigma} -\frac{\i}{2}\mathcal{C}_{IJK}\varepsilon^{\mu\nu\lambda\sigma}
\p_{\sigma}(W^{I}_{\mu}W_{5}^{J}F_{\nu\lambda}{}^{K})\,.
\end{split}
\ee
Using the Bianchi identity and noting that terms like $\varepsilon^{\mu\nu\lambda\sigma}\p_{\mu}\p_\nu$ 
vanish due to symmetry, we find
\be
\eL_{3} \=  \frac{\i}{64}\mathcal{C}_{IJK}t^{I}_{+}\varepsilon^{\mu\nu\sigma\lambda}
\Bigl(12\widetilde{F}^{J}_{\mu\nu}\widetilde{F}^{K}_{\sigma\lambda} - 6t^{I}_{+}
\widetilde{F}^{J}_{\mu\nu}F^{0}_{\sigma\lambda} +t^{J}_{+}t^{K}_{+} F^{0}_{\mu\nu}F^{0}_{\sigma\lambda}\Bigr) +
\eL^{TD}_{VVV}\,,
\ee
where
\be
\begin{split}
\eL^{TD}_{VVV} & \,\equiv\, -\frac{\i \, \varepsilon^{\rho\mu\nu\lambda\sigma}}{8}
\mathcal{C}_{IJK}\p_{\m}\Bigl(2t_{+}A^{J}_{\nu}\widetilde{F}^{K}_{\lambda\sigma} 
- \frac{1}{2}  t_{+}^2 A^{J}_{\nu}\widetilde{F}^{0}_{\lambda\sigma}\Bigr)\,,\\
 & \=  \frac{\i \, \varepsilon^{\rho\mu\nu\lambda\sigma}}{4}\mathcal{C}_{IJK}\p_{\mu}
 \Bigl(W^{I}_{\rho}W^{J}_{\nu}\bigl(2 F^{K}_{\lambda\sigma} +  W^{K}_{\rho}F^{0}_{\lambda\sigma}\bigr) 
+ B_{\nu}W^{I}_{\rho}W^{J}_{\rho}\bigl(W^{K}_{\rho}F^{0}_{\lambda\sigma} - 4F^{K}_{\lambda\sigma}\bigr)\Bigr)\,.
\label{Eqn:LTDVVV}
\end{split}
\ee
This total derivative term is the only source of discrepancy among the 
4d and the 5d Lagrangians as the rest of the terms in~$L_1$ and $L_2$ are gauge-invariant. 
Indeed we can explicitly check that 
\be
\pi \eL_{VVV} - \eL_{4d}^{\text{2-deriv}} \= \frac{1}{8\pi} \eL^{TD}_{VVV}\,.
\label{Eqn:2derivstory}
\ee

\ndt {\bf Higher-derivative Lagrangian}

\vspace{0.2cm}

Now we turn to the four-derivative term~\eqref{Eqn:5dLVWW}. 
The only non-gauge-invariant term is the gravitational Chern-Simons term, $\eL^\text{GCS}$. 
Below, we calculate the total derivative term needed to ensure that the resulting~4d Lagrangian 
is gauge invariant. In the Appendix~\ref{App:rem4d5dfields}, we consider the other terms 
in the higher derivative Lagrangian and explain how they reduce to gauge-invariant terms in~4d.

The gravitational Chern-Simons term is
\be \label{Eqn:L3hd}
\eL^\text{GCS} \= 
-\frac{\i}{128}\varepsilon^{MNPQR}\ic_{I}W_{M}^{I}\bigl(R_{NPAB}(M)R_{QR}{}^{AB}(M) 
+\frac{1}{3}R_{NPj}{}^{i}(V)R^{QR}{}_{i}{}^{j}(V)\bigr)\,.
\ee
We denote the two terms on the right-hand side  by~$\eL^\text{GCS}_1$ (containing the Weyl tensor~$R(M)$) 
and~$\eL^\text{GCS}_2$ (containing the curvature~$R(V)$). 
Using the relation~\eqref{Eqn:R(M)} between $R(M)$ and the Riemann and Ricci tensors, we obtain
\be
\eL^\text{GCS}_1 \= -\frac{\i}{128}\ic_{I}\varepsilon^{MNPQR} W^{I}_{M} \bigl(R_{NP}{}^{AB}R_{QRAB}
-\frac{8}{3}R_{R}{}^{A} R_{NPQA} +\frac{1}{3}R R_{NPQR}  \bigr)\,.
\label{Eqn:L31hd}
\ee
The last two terms in this expression vanish due to symmetry. 
We are therefore just left with the first term, which we expand to obtain
\be
\eL^\text{GCS}_1\= -\frac{\i}{128}\ic_{I}E\varepsilon^{5abcd} W^{I}_{5}R_{ab}{}^{EF}R_{cdEF}
- \frac{\i}{32}\ic_{I} E \varepsilon^{5abcd}W^{I}_{b}R_{5a}{}^{EF}R_{cdEF}\,.
\ee
Using $W^I_{a} = A^I_{a}$, we convert the above expression into two gauge invariant terms and a total derivative term. 
We further simplify these terms using~\eqref{Eqn:RwedgeR} and~\eqref{Eqn:4d5dR5wedge}. After some algebra, we obtain
\be
\eL^\text{GCS}_1 \=  \frac{\i}{256}\ic_{I}t^{I}_{+}\widehat{R}^2_{\wedge} 
- \frac{\i}{32}e \ic_{I}\varepsilon^{abcd}\mathcal{D}_{a}(A^{I}_{b}\widehat{R}_{cd}) 
+\frac{\i}{64}\ic_{I}\widetilde{F}_{ab}^{I} \widetilde{R}_{cd}
\ee
where $\widehat{R}_{\wedge}^2$ is given by~\eqref{Eqn:RwedgeR} and~$\widehat{R}_{cd}$ is given by~\eqref{Eqn:Rtilde}.

The second term in~\eqref{Eqn:L3hd} is
\be
\eL^\text{GCS}_2 \= -\frac{\i}{384}\varepsilon^{MNPQR}\ic_{I} W_{M}^{I} \, R_{NPj}{}^{i}(V) R^{QR}{}_{i}{}^{j}(V)\,.
\ee
To reduce this to~4d, we use~\eqref{Eqn:RVexp} and~\eqref{Eqn:4d5dV}. Several terms vanish due to symmetry and after some simplifying, we find
\be
\varepsilon^{ABCDE}R(V)_{BCji}R(V)_{DE}{}^{ij} \= 4 \varepsilon^{ABCDE}\p_{B}V_{Cij}\p_{D}V_{E}{}^{ji}\,. 
\ee
Using the above equation, we can now write $L^{\text{GCS}}_{2}$ as the sum of a total derivative and gauge invariant term,
\be
\begin{split}
\eL^\text{GCS}_2& \= 
\frac{\i}{768}\varepsilon^{abcd}\ic_{I}t^{I}_{+}\widehat{R}(V)_{abji}\widehat{R}(V)_{cd}{}^{ij} 
+ \frac{\i}{96} \varepsilon^{abcd} \ic_{I} \p_{d} \bigl(\frac{A_{a}^{I}\widehat{R}(V)_{bc}^{ij}Y^{0}_{ij}}{|X^{0}|}\bigr)\\
& \quad \quad - \frac{\i}{192|X^0|}\varepsilon^{abcd} \ic_{I}\widetilde{F}^{I}_{ab} \widehat{R}(V)_{cd}{}^{ij}Y^{0}_{ij}\,.
\end{split}
\ee

Putting the two terms together we can write the Chern-Simons term as the sum of a gauge invariant term 
and a total derivative term,
\be
\begin{split}
\eL^\text{GCS} & \= \frac{\i}{256}\ic_{I}t^{I}_{+}\widehat{R}^2_{\wedge}  
-\frac{\i}{64}\ic_{I}\varepsilon^{abcd}\widetilde{F}_{ab}^{I} \widehat{R}_{cd}
+ \frac{\i \, \varepsilon^{abcd}}{768}\ic_{I}t^{I}_{+}\widehat{R}(V)_{abji}\widehat{R}(V)_{cd}{}^{ij} \\
& \quad \quad\quad \quad- \frac{\i \, \varepsilon^{abcd}}{192|X^0|} 
\ic_{I}\widetilde{F}^{I}_{ab} \widehat{R}(V)_{cd}{}^{ij}Y^{0}_{ij} 
+ \eL^{TD}_{VWW}\,,
\end{split}
\ee
where $\widehat{R}^2_{\wedge}, \,\, \widehat{R}_{cd}$ are given by \eqref{Eqn:RwedgeR} 
and \eqref{Eqn:Rtilde} respectively and the total derivative term is
\be
\eL^{TD}_{VWW} \= \p_{\mu}\Bigl(\frac{\i}{32}\ic_{I}\varepsilon^{\m\n\lambda\sigma}(A^{I}_{\n}\widetilde{R}_{\lambda\sigma}) + 
\frac{\i}{96} \varepsilon^{\m\n\lambda\sigma} \ic_{I} \bigl(\frac{A_{\n}^{I}
\widehat{R}(V)_{\lambda\sigma}^{ij}Y^{0}_{ij}}{|X^{0}|}\bigr)\Bigr)\,.
\label{Eqn:LTDVWW}
\ee

\subsection{Comments on non-F-terms in the 5d/4d reduction \label{sec:nonF}}

We saw above (in \eqref{Eqn:2derivstory}) that, after subtracting the total derivative term \eqref{Eqn:LTDVVV}, 
the two-derivative 5d Lagrangian equals the two-derivative F-term Lagrangian in 4d.
This was, in some sense, inevitable because there is no other supersymmetry- and gauge-invariant
term at two-derivative level. However, at higher-derivative level there are many other invariants,
and there is no reason that the 5d Lagrangian~\eqref{Eqn:5dLVWW}---even after subtracting the total derivative
term~\eqref{Eqn:LTDVWW}---should equal the higher-derivative F-term 4d Lagrangian~\eqref{L4d}. As we see in Section~\ref{sec:Sclass1},
that is indeed not the case.
The discussion of~\cite{Banerjee:2011ts} concentrated only on terms that involved the curvature 
tensors~$R(M), R(V)$ in~5d and in~4d. 
In this subsection we comment on the details of the full Lagrangian and its effect on the BH entropy.

The four-derivative Lagrangian given in~\eqref{L4d} gives rise to four different sets of terms in 4d,
\be \label{Eqn:4derivstory}
8\pi^2\eL_{VWW} - \eL^{TD}_{VWW} = 8\pi(\eL^1_{4d} + \eL^2_{4d} + \eL^3_{4d} + \eL^4_{4d}) \,.
\ee
The first set of terms are F-terms, given by
\be 
\begin{split}
\eL^1_{4d} \= & -\frac{\i}{64}\ic_{I}t^{I}\bigl(2R(M)^{-}_{abcd}R(M)^{-abcd} 
+ \widehat{R}(V)^{-i}_{abj}\widehat{R}(V)^{-abj}{}_{i}\bigr)\\
& -\frac{\i}{512}T^{-ab}(X^0)^{-1}\ic_{I}(Y^{ijI}-t^I Y^{ij0})\widehat{R}(V)^{-k}_{abj}\varepsilon_{ki}\\
& + \frac{\i}{256}\ic_{I}(X^0)^{-1}T^{-}_{cd}\widehat{R}(M)^{abcd}(\widehat{F}^{-I}_{ab}-t^I F^{-0}_{ab}) + h.c.\,.
\end{split}
\ee
The sum of this term and the second derivative Lagrangian discussed in the previous section leads to form the~4d F-term 
Lagrangian~\eqref{L4d}, including the higher derivative coupling.
The second set of terms are of the Gauss-Bonnet type,
\be
\begin{split}
\eL^2_{4d}& \= -\frac{\i}{384}\ic_{I}t^{I}\bigl(\frac{2}{3}R_{ab}R^{ab} 
+ R^{+i}_{abj}(\widehat{V})R^{+abj}{}_{i}(\widehat{V})\bigr)\\
&\quad \quad -\frac{\i}{768 X^0}\ic_{I}t^{I}_{-}T^{-}_{cd}R^{abcd}(M)F^{-0}_{ab} + h.c.\, .
\end{split}
\ee
The emergence of these terms led to new~$\CN = 2$ 
supersymmetric invariants in~4d,~\cite{Butter:2013lta}.
The third set of terms are D-type terms (full-superspace integrals), 
\be
\begin{split}
eL^3_{4d} & \= \frac{\i}{384|X^0|^2}\ic_{I}R^{+ab}_{ij}(\widehat{V})\bigl(F^{+I}_{ab}Y^{ij0} 
- F^{+0}_{ab}Y^{ijI} + t^{I}_{-}F^{+0}_{ab}Y^{ij0}\bigr)\\
& \quad \quad + \frac{\i}{1536 X^0}T^{-ab}\ic_{I}
\bigl(Y^{ijI} -(t^{I}_{-} + t^I)Y^{ij0}\bigr)R^{+}_{abij}(\widehat{V}) + h.c.\,.
\end{split}
\ee
The fourth set of terms are kinetic terms for the auxiliary field~$T$ as well as quartic terms in the scalar 
field~$X$ and the field strengths~$F^0$. They also include coupling between the square of the derivatives 
of scalar fields, field strengths and the gauge fields. We leave the details of these terms to future work.
One possibility is that we need to consider supersymmetric couplings coming from   
supersymmetrization of the square of the Ricci scalar discussed in~\cite{Ozkan:2013nwa,Baggio:2014hua}.

In the next section we discuss the effect of these total derivative terms on the black hole entropy. 
One interesting observation regards the value of these four sets of terms on the black hole solution.
The F-type terms evaluate to give the 4d entropy. 
The D-terms vanish, as expected from the result of~\cite{Murthy:2013xpa}. 
Somewhat surprisingly the Gauss-Bonnet term and the fourth set of terms are both non-zero, 
but they exactly cancel each other. The total value of the action is consistent with the fact that 
the 4d black hole entropy is completely recovered by the F-type terms,~\cite{deWit:2010za, Murthy:2013xpa}.

\section{The entropy function and charges in 4d and 5d \label{sec:Sclass}}

In this section, we discuss the quantum entropy formalism as applied to the BMPV black hole. 
As we review below, the main objective is to calculate a functional integral of a supersymmetric 
Wilson line with the supersymmetric off-shell action. As mentioned in Section~\ref{sec:action}, 
there is an important subtlety that the action in 5d is not locally gauge invariant because of the 
presence of Chern Simons terms. This creates an ambiguity in the definition of the Wilson line 
(with regards to the notion of charge) and the action that is to be used in the functional integral. 
In order to resolve this ambiguity we will use the 4d/5d connection. In Section~\ref{sec:action}, 
we discussed the 4d/5d connection in the context of the off-shell action. This discussion is 
particularly useful when combined with the idea of the black hole  
4d/5d connection of~\cite{Katz:1999xq},~\cite{Gaiotto:2005gf}.
In this context, the 4d supersymmetric BH is thought of as a 5d supersymmetric BH
placed at the origin of Taub-NUT space. This implies that the near-horizon configurations
of the four-dimensional and the five-dimensional systems are exactly the same. It follows 
that the quantum entropy (which is defined on the near-horizon configuration) of these 
two BHs must be equal. This point of view clarifies the correct notion of charge and action 
that should be used for the quantum entropy. This point of view has been productively 
used in \cite{Castro:2007ci}. We will lift this point of view to the problem of the quantum entropy.

As discussed in~\cite{Gupta:2019xac, deWit:2009de,Gomes:2013cca} the near-horizon Euclidean
metric of the black hole is 
\be \label{Eqn:AdS2S2S1euc}
\begin{split}
ds^{2} & \= \sinh^{2} \eta d\theta^{2} + d\eta^{2} + d\psi^{2} + \sin^{2}\psi d\phi^{2} + \cosh^{2}\alpha\bigl(d\rho + B\bigr)^{2}\,, \\
B & \= + \cos\psi d\phi - \tanh\alpha(\cosh\eta - 1)d\theta\,.
\end{split}
\ee
We have chosen the parameter~$\a$ such that it is real in the Euclidean theory. 
Recall that in the Lorentzian theory the metric and the gauge field~$B$ are real, which 
therefore~$\a$ is imaginary.

The coordinates have the following ranges
\be
\eta \in [0,\infty]\,,\quad \quad \theta\,, \phi \in [0, 2\pi]\,, \quad \quad \psi\in [0, \pi]\,,\quad\quad\rho \in [0, 4\pi]\,.
\ee
The near horizon configuration of the fields $D, T$ is
\be \label{Eqn:NHaux}
D \= 0\,, \quad
T_{\theta\eta} \= -\frac{\i}{4}\sinh\eta\cosh\alpha\,, 
\quad T_{\psi\phi} \= \frac{\i}{4}\sin\psi\sinh\alpha\,.
\ee
For the vector multiplet fields, the near horizon supersymmetry requires that the value of the scalar field is constant~i.e.,
\be\label{Eqn:AttractorSigma}
\sigma^{I}=\sigma^{I}_{*}\,,
\ee
with the constants~$\sigma^{I}_{*}$ are determined by the charges of the black hole~$q_{I}$. 
The charges of the black hole are $q_{I}\,, J$.
The BPS equations for the field strength is 
\be\label{Eqn:NearHorzFieldSt}
F_{*}{}^{I}_{AB} \= 4\sigma_{*}^{I}\,T_{AB}\,.
\ee
For the gauge field, we integrate to obtain
\be \label{Eqn:NHGauge}
W_{*} \=  \i\, \sigma_{*}^{I}\cosh\alpha\, ( \cosh\eta-1) \, d\theta 
- \i\, \sigma_{*}^{I}\sinh\alpha \cos\psi \, d\phi - \i \, \sigma_{*}^{I}\sinh\alpha\, d\rho\,.
\ee
Now we recall 4d quantum entropy function.

\subsection{Elements of the quantum entropy function \label{EleQEF}}
The quantum entropy of a supersymmetric black hole was defined in~\cite{Sen:2008vm} to be
the expectation value of the Wilson line,
\be \label{qefdef}
Z(\vec{q})\; \equiv \; \text{e}^{\mathcal{S}^{\text{qu}(\vec{q})}} \defeq  \big{\langle} 
\exp \bigl( S_\text{WL} \bigr) \big{\rangle}_\text{AdS$_2$}^\text{finite}
\ee
The subscript~``AdS$_2$" means that we integrate over 
all asymptotic Euclidean AdS$_2$ field configurations, and the superscript ``finite" refers 
to the fact that we renormalize the action by subtracting the divergences coming from the 
infinite volume of~AdS$_2$, according to holographic renormalization~\cite{Skenderis:2002wp}.
The Wilson line in 4d is given by\footnote{We have used the quantization of charge in this equation 
as~\cite{Dabholkar:2010uh}, which may differ by a factor of two with other quantization used 
in the literature as~\cite{Mandal:2010cj}.}
\be \label{Eqn:S4dWL}
S_{W.L} \= \i\, \frac{q^{4d}_{\widetilde{I}}}{2} \, \int\limits_{0}^{2\pi} A^{\widetilde{I}}_{\theta} \, d\theta\,.
\ee

A method to calculate this functional integral using localization in supergravity was given 
in~\cite{Dabholkar:2010uh}. In that paper, this method was applied to 
supersymmetric black holes in asymptotically flat four-dimensional space, 
wherein the fully supersymmetric on-shell configuration is~AdS$_2 \times$~S$^2$, with the  
symmetry algebra~$SL(2) \times SU(2)$. 
The~$SL(2)$ algebra with generators~$L_0, L_{\pm}$ acts on the~AdS$_2$ space,
and the~$SU(2)$ algebra with generators~$J_0, J_{\pm}$ acts on the~AdS$_2$ space,
The method relies on the existence of a complex supercharge which obeys the 
off-shell algebra~$\CQ^2=L_0-J_0$. One deforms the action by the term~$ \lambda \CQ \CV$ 
with~$\lambda >0$ a real parameter, which is chosen to satisfy~$\CQ^2 \CV =0$. 
The functional integral turns out to be independent of~$\lambda$ and it is easy to calculate 
it at the value~$\lambda \to \infty$. 
The resulting formula which is exact in perturbation theory is an integral over the set of field configurations 
annihilated by the localizing supercharge. The integrand is the renormalized action of the off-shell 
supergravity, and one includes the 1-loop determinant of the deformation action~$\CQ \CV$ over the non-BPS 
directions. 

The result is that the quantum entropy function can be reduced to
\be \label{Eqn:Zqu}
Z(\vec{\widehat{q}})\= e^{\mathcal{S}^{\text{qu}(\vec{\widehat{q}})}} \= \int_{\mathcal{M}} \, 
e^{S_{\text{Bulk}}^{\text{Ren}}  + S_{\text{W.L}}^{\text{Ren}}(\vec{\widehat{q}})} \,Z_{\text{1-loop}}\,,
\ee
where $\vec{\widehat{q}}$ are the charges parameterising the black hole. 
The integral is evaluated over the localization manifold,~$\mathcal{M}$. 
Here the renormalised bulk action is 
\be
S_{\text{Bulk}}^{\text{Ren}} \= S_{\text{Bulk}} + S^{1(4d)}_{\text{BND}}\,,
\ee
where~$S_{\text{Bulk}}$ is given by evaluating~\eqref{L4dpieces} on the localization manifold, 
and the counter-term is given by 
\be \label{Eqn:SbulkBnd}
S^{1(4d)}_{\text{BND}}  \= -\i \,\int\limits_{0}^{2\pi}(F(X) - \overline{F(X)}) \,e^{\theta}_{\theta} \, d\theta \,,
\ee
The Wilson line~\eqref{Eqn:S4dWL} is not supersymmetric, however, it can be 
made supersymmetric by adding the following term
\be \label{Eqn:WLBnd}
S_{W.L}^{\text{Ren}} \= S_{W.L} + S^{2(4d)}_{\text{BND}}\,,
\ee
where the precise definition of this counter-term follows from demanding supersymmetry
\be
S^{2(4d)}_{\text{BND}} \= -\int\limits_{0}^{2\pi} \frac{q^{4d}_{\widetilde{I}}}{2}(X^{\widetilde{I}} 
+ \overline{X}^{\widetilde{I}})\, e^{\theta}_{\theta} d\theta\,.
\ee
One can see that~$S_{W.L}^{\text{Ren}}$ is supersymmetric by computing the variations 
under supersymmetry of each of the terms and seeing that they cancel each other exactly. 
This is the Maldacena Wilson loop first found in~\cite{Maldacena:1998im}.

In~\cite{Dabholkar:2010uh, Gupta:2012cy}, the localization manifold for the 4d black hole was found. 
The off-shell treatment allows one to separate the problem into two parts, finding the localization manifold 
of the Weyl multiplet and finding the localization manifold of the vector multiplet. The Weyl multiplet was 
found to be fixed to the attractor values of the black hole and the non-zero vector multiplet was found to be given by
\begin{equation} \label{4dlocsols}
X^{\widehat{I}} \= X^{\widehat{I}}_{*} + \frac{C^{\widehat{I}}_{4d}}{\cosh\eta}\,, \quad
Y^{\widehat{I}12}_{4d} \= \frac{C^{\widehat{I}}_{4d}}{\cosh^2 \eta} \,,
\end{equation}
where the attractor value of the scalar field is
\be
X^{\widehat{I}}_{*} \= \frac{e^{\widehat{I}}_{*} + \i \, p^{\widehat{I}}_{*}}{2}\,,
\ee
where $e^{\widehat{I}}_{*}\,, p^{\widehat{I}}_{*}$ are the electric and magnetic charges of the black hole.

Over the localization manifold, the renormalised action appearing in the quantum entropy function in 4d is
\be \label{Eqn:stotal4d}
S_{\text{Ren}} \= - \pi q_I \v^I +  \CF (\v,p)\,,
\ee
where 
\be 
 \CF (\v,p) \= - 2 \pi i \Bigl(F(\frac{\v^I+ \i \, p^I}{2}) - \overline{F}(\frac{\v^I- \i \, p^I}{2}) \Bigr)\,. 
\ee
The renormalised action is the Legendre transform of the bulk renormalised action. 
As discussed in~\cite{Sen:2008}, using a saddle point approximation, the semi-classical 
entropy comes from evaluating $S^{\text{Ren}}$ at its saddle points. That is
\be \label{Eqn:GSCE}
\mathcal{S}^{\text{BH}}(\vec{\widehat{q}}) \= S^{\text{Ren}}_{*}(\vec{\widehat{q}})+\dots\,,
\ee
where $S^{\text{Ren}}_{*}$ is the renormalised action evaluated at the saddle points. 
Semi-classically solving for the entropy is reduced to an extremization problem.
We wish to consider a similar analysis for the entropy of the 5d black hole using the 4d/5d connection. 
There is an ambiguity in the notion of charge as discussed above due to the presence of the 
Chern-Simons term. In the next sections, we will review the different notions of charge.

\subsection{Maxwell and Page charges}

An important question in the context of black hole entropy is: what is the charge carried by the black hole? 
It has a clear answer in 4d, since there is a unique definition of electric charge. In contrast to it, in 5d, 
there is no unique definition of charge, and hence the entropy depends on which definition of charge 
is used. This ambiguity in the definition of electric charge is a direct consequence of the presence 
of the Chern Simons term in the Lagrangian, which is not gauge invariant. Two common notions 
of electric charge which are used are the Page charge and Maxwell charge~\cite{Marolf:2000cb}.
The Page charge is conserved but not gauge invariant, while the Maxwell charge is gauge invariant 
and conserved. The definition of charge that we should use depends on the physical question under consideration.

To present the discussion of charges in general, we consider the action of a single vector field given by 
\be
S \=\int d^{5}x\sqrt{g}\,\Big[R_{5}+\frac{c_{0}}{4}F_{\mu\nu}F^{\mu\nu}+c_{1}
\varepsilon^{\mu\nu\rho\sigma\delta}W_{\mu}F_{\nu\rho}F_{\sigma\delta}
+c_{2}\varepsilon^{\mu\nu\rho\sigma\delta}W_{\mu}R_{ab\nu\rho}R_{ab\sigma\delta}+...\Big]\,,
\ee
where $c_{0},c_{1}$ and $c_{2}$ are some constants and dots represent terms independent 
of~$F_{\mu\nu}$ and~$W_{\mu}$. One can relate the above action with the Lagrangian 
given in~\eqref{Eqn:5dLVVV} and~\eqref{Eqn:5dLVWW} for a single vector multiplet by 
replacing the auxiliary fields by their on-shell value and $\sigma$ to be constant. 
The equation of motion of the gauge field is 
\be
dE\=0\,,
\ee
where $E$ is a 3-form whose components are given by
\be\label{eqnE}
E_{\m\n\g}\=\ve_{\m\n\g\alpha\beta}(c_{0}F^{\alpha \beta}+6\, c_{1}\ve^{\alpha\beta\kappa\eta\chi}
W_{\kappa}F_{\eta\chi} +2\,c_{2} \ve^{\alpha\beta\kappa\eta\chi}(\omega^{ab}_{\kappa}\p_{\eta}\omega^{ab}_{\chi}
-\frac{2}{3}\omega^{ab}_{\kappa}\omega^{af}_{\eta}\omega^{fb}_{\chi}))\,.
\ee

The Page charge is the charge obtained by the Gauss law constraint, 
\be
Q_{5}\=\int_{\Sigma^{\infty}_{3}} E\,,
\label{Eqn:Page}
\ee
where $\Sigma^{\infty}_{3}$ is a 3-surface at spatial infinity. As it is evident from~\eqref{eqnE}, 
that the Page charge is gauge dependent. However, this non-invariance under the gauge 
transformation, both local and large, disappears if the gauge field falls off sufficiently fast 
at asymptotic infinity.  An important property of the Page charge is that it is localized in 
the sense that in the absence of any sources, the Page charge vanishes. It implies that 
the Page charge is independent of the 3-surface i.e. 
\be
\int_{\Sigma^{\infty}_{3}} E\=\int_{\Sigma_{3}} E\,.
\ee
where $\Sigma_{3}$ is any arbitrary 3-surface surrounding the black hole. 
Thus, the Page charge calculated at asymptotic infinity is the same as the charge 
computed near the horizon.

Next, we define the Maxwell charge. The Maxwell charge is defined as 
\be
\wt  Q_{5}\=c_{0}\int_{\Sigma^{\infty}_{3}} *F\,.
\ee
It is clearly gauge invariant under both local and global gauge transformations. 
As it is evident from the definition, the Maxwell charge differs from the Page charge 
as the former does not have the contribution from the Chern Simons terms. In this case, 
we see from the equation of motion~\eqref{eqnE} that both the gauge field and spin 
connection are sources for the Maxwell charge. As a result, the Maxwell charge is 
not localized and, therefore is not independent of the deformation of the 
3-surface $\Sigma^{\infty}_{3}$\footnote{Although the Maxwell charge depends on 
the choice of the 3-surface, it is interesting to compare the Maxwell charge computed 
near the horizon with the Page charge. Using the gauge field configuration near the 
horizon~\eqref{Eqn:NHGauge}, the difference at the 2-derivative order is
\be
Q_{5}-\wt Q_{5}=12 c_{1}\int d\psi\,d\phi\,d\rho W_{\rho}F_{\psi\phi}
=12c_{1}\sigma^{2}_{*}\int d\psi\,d\phi\,d\rho\sinh^{2}\alpha\sin\psi\,.
\ee
Thus, we see that the Maxwell charge coincides with the Page charge for a 
non-rotating black hole for which $\sinh\alpha=0$. However, the difference arises 
when we incorporate the angular momentum and higher derivative terms.}.  

Which of the above charges is useful for the black hole entropy computation? 
Let's focus on the semi-classical limit in which the problem reduces to solving an 
extremization equation as discussed in~\eqref{Eqn:GSCE}. From that equation, it is clear that 
extremization with respect to the gauge field involves the whole Lagrangian (as in 
the Gauss law constraint), not just quadratic terms~\cite{Sen:2008}.
Thus, according to the above discussion it would seem that the Page charge is the 
correct charge to describe the black hole entropy. Indeed, we find that taking the Page 
charge reproduces the correct entropy at at the 2-derivative order. However, as we will 
see below this does not work at higher derivative order. The reason is that the classical 
entropy function requires gauge invariance along the~AdS$_{2}$ direction, which leads 
us to reduce the original theory to a gauge-invariant theory in 4d. As a result, the electric 
charge following from the gauge-invariant entropy function differs from the Page charge. 
The difference between the two charges is equal to the contribution from total derivative 
terms one adds to reduce to 4d gauge-invariant theory.

\subsection{Chern-Simons term in Sen's formalism}
As we saw in the previous section the entropy of the black hole is completely fixed 
in terms of near-horizon data. The attractor mechanics for extremal black holes fixes 
the scalar fields at the horizon in terms of near-horizon charges. Thus, the macroscopic 
entropy is given in terms of near-horizon charges. Note that these charges, in general, 
are different from the asymptotic charges, which also includes the contribution from the 
hair modes that has support outside of the 
horizon~\cite{Castro:2008ys,Banerjee:2009uk},~\cite{Dabholkar:2010rm}. 

Sen's (classical) entropy function relates the near-horizon charges with the near-horizon 
value of the scalar fields. The classical entropy function approach is based on the 
observation that the near-horizon geometry has the form AdS$_{2}$ times a compact 
space and the solution of the equation of motion respect the isometry of the background. 
This approach to computing the entropy of an extremal black hole follows directly from 
the extremal limit of the Wald entropy formula and is based on the assumption of the 
Lagrangian density is diffeomorphic and gauge invariant. 
The situation is subtle when the theory of gravity described by the Lagrangian is not 
diffeomorphic and/or gauge invariant, for example when the Lagrangian has 
Chern-Simons terms. A Chern-Simons term, which is not locally gauge invariant, 
explicitly violates the assumption used in Wald's formalism. 

It is also important to note that in the classical entropy function, the electric charge 
is defined by varying the Lagrangian with respect to~$F_{\theta\eta}$. This is not 
correct when there is explicit gauge Chern-Simons term. Thus, the entropy function 
approach can not be applied directly for the Lagrangian having Chern-Simons term. 
However, it is still possible to compute the entropy of an extremal black hole in such 
theory using the classical entropy function. This has to do with the fact that on 
dimensional reduction, a Chern-Simons term (gauge or gravitational or mixed) 
reduces to a gauge-invariant term and a total derivative. As a result, the original 
Lagrangian together with an appropriate boundary term on the dimensional reduction 
reduces to a Lagrangian which is locally gauge invariant in the two-dimensional 
$AdS_{2}$ space. As a result, the reduced Lagrangian does not contain any explicit 
connection term, and all the terms are a function of field strength/curvature.
This is the Sen's proposal for dealing with Chern-Simons terms in the classical entropy function.

In the present case of the rotating black hole in 5d, we have a similar situation. 
The near-horizon geometry has the form AdS$_{2}\times$S$^{2}\rtimes$S$^{1}$. 
Therefore, to compute the entropy using classical entropy function one reduces the 
theory to 4d (or to~AdS$_2 \times$~$S^{2}$). To begin with, the Lagrangian has two 
kinds of Chern-Simons terms. One is the gauge Chern-Simons term that appears at 
the 2-derivative order \eqref{Eqn:L1}, and another is the mixed gauge-gravitational 
Chern-Simons term that appears at the 4-derivative order~\eqref{Eqn:L3hd}. The 
Lagrangian, together with total derivative terms~\eqref{Eqn:LTDVVV} 
and~\eqref{Eqn:LTDVWW}, reduces to a gauge-invariant Lagrangian in 4d.
One can now apply the entropy function on the 4d Lagrangian~\cite{Castro:2007ci}. 
We will follow this strategy for the entropy computation in the off-shell set up. 
In particular, we will find the attractor equation~\eqref{Eqn:fulleqSP} for our black hole, 
which agrees with the attractor equation in 4d.

Finally, it also brings us to an important difference with the result of~\cite{deWit:2009de}. 
In that paper, the authors found that the 4d and 5d entropies are not the same at the 
higher derivative order. They pointed out that the discrepancy is due to a difference 
between their attractor equation and the 4d one. 
The authors found the attractor equation using the Gauss law constraint, which is 
essentially the Page charge computation, and is given by
\be\label{5d:AttractorEq}
q_{I} \= 6\,C_{IJK}\sigma_{*}^{J}\sigma_{*}^{K}-\frac{3}{8}c_{I}\cosh^{2}\alpha\,.
\ee
As we will see below this is different from the 4d attractor equation~\eqref{Eqn:fulleqSP}. 
At the 2-derivative order our attractor equation agrees with \eqref{5d:AttractorEq} 
with~$\ic_{I} = 0$, but as we will see even that is accidental. 

\subsection{Supersymmetric Wilson lines and bulk action in 5d}
As discussed above the definition of charge to take in Sen's formalism is found 
by reducing to gauge invariant 4d theory, where the definition of charge is unambiguous.
Using the $4d/5d$ connection described in section~\ref{sec:4d5dfields}, the 
Wilson line lifts, \eqref{Eqn:S4dWL} to
\be \label{Eqn:SWL}
S_{W.L} \= \frac{\i \, J}{2}\int\limits_{0}^{2\pi}B_{\theta} d\theta 
+\frac{ \i \, q_{I}}{2}\int\limits_{0}^{2\pi} (W^{I}_{\theta} - W^{I}_{\rho} B_{\theta} )d\theta\,.
\ee
where $J$ is the angular momentum of the black hole and $q_{I}$ are the 
charges of the black holes. The 5d Wilson line is
\be \label{Eqn:SWL5d}
S^{5}_{W.L.} \= S_{W.L.} - \Delta S_{W.L.}  
\= \frac{\i \, q_{I}}{2}\int\limits_{0}^{2\pi}W^{I}_{\theta} d\theta\,,
\ee
where the difference is
\be
\Delta S_{W.L.} \= \frac{\i \,J}{2}\int^{2\pi}_{0}B_{\theta}d\theta -\frac{\i \,q_{I}}{2}\int^{2\pi}_{0} 
B_{\theta} W_{\rho}^{I}d\theta\,.
\ee

The boundary terms in 5d are lifted from the~4d boundary terms as
\be \label{Eqn:Lbnd}
S^{1}_{\text{BND}} \= 4\, S^{1(4d)}_{\text{BND}}\,, \quad \quad S^{2}_{\text{BND}} 
\= 4\, S^{2(4d)}_{\text{BND}}\,.
\ee
where the factor of $4$ arises from the integral over the fifth direction 
(and different normalizations in the 4d/5d Lagrangians~\eqref{Eqn:2derivstory}) 
as discussed in~\eqref{Eqn:Factorof4}.

On the near-horizon solution, the action~\eqref{Eqn:5dLVVV}-\eqref{Eqn:5dLVWW} evaluates to
\be
\begin{split}
S_{VVV} + S_{VWW} & \= 2\pi \, \CC(\sigma_{*})\cosh\alpha(\cosh\eta_{0} - 1) 
-\frac{3\pi}{8}\ic_{I}\sigma^{I}_{*}\cosh\alpha(\cosh\eta_{0} -1) \\
&\qquad -\frac{3\pi}{16}\ic_{I} \, \sigma^{I}_{*}\sinh^2 \alpha\cosh\alpha(\cosh\eta_0 - 1) \,,
\label{Eqn:Sbulk}
\end{split}
\ee
where we have introduced a large cut-off $\eta_{0}\,$. 
On the classical solution the total derivative term evaluates to
\be
\begin{split}
S^{TD}_{VVV} + S^{TD}_{VWW}& \=  
2\pi \, \CC(\sigma_{*})\frac{\sinh^2\alpha}{\cosh\alpha}(\cosh\eta_0 - 1)
+ \frac{\pi}{8}\ic_{I}\sigma^{I}_{*}\cosh\alpha(\cosh\eta_{0} -1)\\
&\qquad -\frac{3\pi}{16}\ic_{I}\sigma^{I}_{*}\sinh^2 \alpha\cosh\alpha(\cosh\eta_0 - 1)\,.
\label{Eqn:STD}
\end{split}
\ee
In the next section, we discuss how to define the renormalised action appearing 
in the quantum entropy function for the $5d$ black hole.

\subsubsection*{5d attractor equations at 2-Derivative level}

At two derivative level for a static black hole, the attractor equation for Page 
charge on our black hole,~\eqref{5d:AttractorEq}, can be obtained by evaluating 
the sum of the action~\eqref{Eqn:Sbulk} and the Wilson line~\eqref{Eqn:SWL} 
at the saddle points.
The saddle point equations are given by setting
\be
\frac{\p S}{\p \v^{I}_{*}}   \= 0\,, \quad \quad 
\frac{\p S}{\p \v^{0}_*} \=0\,,
\ee
where the variation is with respect to the conjugate variables~$\v^I$ and~$\v^0$ to the charges~$q_{I}$ 
and angular momentum~$J$ in the Legendre transform.
In the static case, the total derivative term evaluates to zero. In addition as the metric 
is not fibred over $AdS_{2}$, the 4d and 5d Wilson lines also coincide. 

For a spinning black hole, as discussed above, Sen's formalism suggests we reduce 
our action to 4d and write use the gauge invariance to define the action.
On the black hole using~\eqref{Eqn:AdS2S2S1euc} and~\eqref{Eqn:NHGauge}, the 
lifted Wilson line evaluates to
\be
S^{\text{Ren}}_{W.L.}  \= 
\i\, J\pi \tanh\alpha +\frac{\pi q_{I} \sigma^I_{*}}{\cosh\alpha}  \,,
\ee
where we have renormalised the Wilson line by adding $S^2_{\text{BND}}$.
The attractor equation relating the charge of the black hole and the scalar fields, 
is the saddle point equation of the sum of the renormalised action and Wilson line. 
\be 
S^{\text{Ren}}_{VVV}-S^{\text{Ren} T.D.}_{VVV} + S^{\text{Ren}}_{W.L.} \=
 \pi q_{I}\varphi^{I}_{*} + \pi J \varphi^{0}_{*}-\frac{2\pi \CC(\varphi_{*})}{1+(\varphi^{0}_{*})^{2}}\,,
\ee
where we introduce new variables that are conjugate to the angular momemtum~$J$ and charges~$q_{I}$,
\be \label{Eqn:Newvarstar}
\varphi^{0}_{*}  \= \i\,\tanh\alpha \,, \quad \quad
\varphi^{I}_{*} \= \frac{\sigma^{I}_{*}}{\cosh\alpha}\,.
\ee
Varying with respect to the conjugate variables and rewriting in terms of the scalar field~$\sigma$, we obtain
\be
q_{I} \= 6 \, \CC_{IJK}\sigma^J_{*}\sigma^K_{*}\,,
\qquad
J \= -4 \, \i \, \CC(\sigma_{*})\sinh\alpha
\label{Eqn:qI2der}
\ee

As mentioned above, a coincidence occurs at the two derivative level. If we do not include the 
total derivative term or modify the Wilson line, we obtain the same equation for the charge.
\be
S^{\text{Ren}}_{VVV}+S^{\text{Ren} 5}_{W.L.} \= \pi q_{I} \widehat{\varphi}^{I}_{*} 
- \frac{2\pi \CC(\widehat{\varphi}_{*})}{\cosh^2\alpha}\,,
\ee
where the conjugate variable to $q_{I}$ is $\widehat{\varphi}^I_{*} = \sigma_{*}^I \cosh\alpha$. 
We vary with respect to the conjugate variable to get the extremization equation. Rearranging 
and rewriting in terms of the scalar field $\sigma$, we arrive back at the equation for $q_{I}$ 
in~\eqref{Eqn:qI2der}. This accident occurs because for the attractor equation~\eqref{Eqn:qI2der}, 
the difference in the 4d and 5d Wilson line is exactly equal to the total derivative term we add to the action.
\be
\Delta S_{W.L.} \= -\frac{2\pi\CC(\sigma_{*})\sinh^2 \alpha}{\cosh\alpha} \= S_{VVV}^{T.D}
\ee
This accident does not occur at higher derivative level.

\subsubsection*{5d attractor equations including Higher Derivative}

Including higher derivative terms, the action is
\be
S_{\text{Ren}}(q_{I}, J; \varphi^{I}_{*}, \varphi^{0}_{*}) 
\=  \pi q_{I}\varphi^{I}_{*} + \pi J \varphi^{0}_{*}-\frac{2\pi \CC(\varphi_{*})}{1+(\varphi^{0}_{*})^{2}} + \frac{\pi}{2} \frac{\ic_{I}\varphi^{I}_{*}}{1+(\varphi^{0}_{*})^{2}}\,,
\label{Eqn:stotalstar}
\ee
in terms of the conjugate variables defined in~\eqref{Eqn:Newvarstar}.
The saddle point equations now take the form 
\be
\label{Eqn:fulleqSP}
q_{I} \= 6\CC_{IJK}\sigma^{J}_{*}\sigma^{K}_{*} - \frac{1}{2}\ic_{I} \cosh^2 \alpha\,, \quad 
J \= -4\,\i \,\CC(\sigma_{*})\sinh\alpha + \i\, \pi \ic_{I}\sigma_{*}^{I} \sinh\alpha \cosh^2 \alpha\,,
\ee
The attractor equation for the charge is different to the Page charge~\eqref{5d:AttractorEq}, 
even in the case of a static black hole.

For the static black hole we repeat the above process without including total derivatives and 
using the 5d Wilson line,
\be
S^{\text{Ren}}_{VVV}+S^{\text{Ren}}_{VWW} + S^{\text{Ren}5}_{W.L} \= \pi q_{I}
\widehat{\sigma}_{*}^I  - 2\pi \CC(\widehat{\sigma}_{*}) + \frac{3\pi}{8}\ic_{I}\sigma_{*}^I\,,
\ee
we find the saddle point equation for $q_{I}$ is
\be
q_{I} \= 6\,\CC_{IJK}\sigma^{J}_{*}\sigma^{K}_{*} - \frac{3}{8}\ic_{I}\,.
\ee
This coincides with the Page charge~\eqref{5d:AttractorEq} but does not agree with the 
charge~\eqref{Eqn:fulleqSP}. As was discussed in the above section, the correct definition 
of charge for the black hole entropy is to use Sen's formalism and reduce to a lower 
dimension in which the charge is unambiguous. Using the Page charge,~\cite{Banerjee:2011ts} 
found that the entropy of the 5d black hole does not agree with the entropy of the 4d black hole. 
This is puzzling because microscopically these black holes are the same as discussed at the start of this section. 
As we show below, if we define charge by include the total derivative terms and the 
Wilson line~\eqref{Eqn:SWL}, we get the same entropy for the black holes.

\subsection{Semiclassical entropy \label{sec:Sclass1}}

To calculate the semi-classical entropy of the spinning black hole, we start with the 
action~\eqref{Eqn:stotalstar}. We write the saddle point equations~\eqref{Eqn:fulleqSP} 
in terms of the conjugate variables,
\be
\begin{split} \label{Eqn:Saddles}
q_{I} & \= \frac{6\, \CC_{IJK} \, \varphi^{J}_*\varphi^{K}_*}{1+(\varphi^{0}_*)^2}-\frac{\ic_{I}}{2(1+(\varphi^{0}_*)^2)}\,,\\
J & \= -\frac{4 \, \CC(\varphi_{*})\, \varphi_{*}^{0}}{(1+ (\varphi^{0}_*)^2)^2} 
+ \frac{\ic_{I} \, \varphi_{*}^{I} \, \varphi_{*}^{0}}{(1+(\varphi^{0}_*)^2)^2}\,.
\end{split}
\ee
The semi-classical entropy is given by evaluating the action at the saddle points. 
In order to do this, we solve the above equations for the conjugate variables in terms 
of the charges of the black hole.
If we exclude higher derivative dependence, we can solve the above equations exactly. We find
\be
\begin{split}
\varphi^{0}_* & \= -\frac{J}{2\sqrt{Q^3-\frac{1}{4}J^{2}}} + \mathcal{O}(\ic_{I})\,,\\
\varphi^{I}_* &\= \widehat{q}^{I} \sqrt{\frac{Q^3}{Q^3 - \frac{1}{4}J^{2}}}+ \mathcal{O}(\ic_{I})\,, 
\end{split}
\ee
where 
\be 
\widehat{q}^{I} = \frac{\varphi^{I}_*}{\sqrt{1+(\varphi^0_*)^2}}\,,
\ee
and $\widehat{q}^{I}$ is implicitly defined by
\be
q_{I} \= 6 \, \CC_{IJK} \widehat{q}^{J}\widehat{q}^{K}\,,
\label{qiqihatrel}
\ee
and
\be
Q^{\frac{3}{2}} \defeq 2 \, \CC(\widehat{q})\,.
\ee
Substituting this into \eqref{Eqn:stotalstar} we get
\be
S^{BH}_{*}(Q, J) \= 2\pi \, \sqrt{Q^3 - \frac{1}{4}J^2}+ \mathcal{O}(\ic_{I})\,. 
\label{SBHQ32}
\ee

Including higher derivatives, we can solve\eqref{Eqn:Saddles} perturbatively 
by expanding about the two derivative solution.
We find to first order in $\ic_{I}$, that the fields are
\be
\begin{split}
\varphi^{0}_* & \= -\frac{J}{2\sqrt{Q^3-\frac{1}{4}J^{2}}} 
- \frac{\ic_{I} \, \widehat{q}^{I}J}{8 \, Q^{\frac{3}{2}}\sqrt{Q^3 - \frac{1}{4}J^2}} + \mathcal{O}(\ic_{I}^2)\,,\\
\varphi^{I}_* &\= \widehat{q}_{I} \, \sqrt{\frac{Q^3}{Q^3 - \frac{1}{4}J^{2}}} 
+ \frac{\ic_{J} \, \CC^{IJ}(Q^3 - \frac{1}{4}J^2)}{24 \, Q^3} + \mathcal{O}(\ic_{I}^2)\,. 
\end{split}
\ee
Here we define $\CC^{IJ}$ as the inverse of~$\CC_{IJ} \equiv \CC_{IJK}\widehat{q}^K$.
Substituting these back into the action, we get the semi-classical entropy with higher derivative correction to be
\be
\eS^{BH}_{*}(Q, J) \= 2\pi \, \sqrt{Q^3 - \frac{1}{4}J^2} \; \Bigl(1+ \frac{\ic_{I}
\widehat{q}^{I}}{4\,Q^{\frac{3}{2}}}\Bigr)+\mathcal O(c_{I}^{2})\,.
\ee
The above formula is in agreement with the entropy of the black hole in~$4d$. 
We now proceed to discuss the quantum entropy.

\section{The localization manifold and its off-shell action \label{sec:locaction}}

In Section~\ref{EleQEF} we presented the general elements of the quantum entropy formula. 
In the present section we discuss the details of the localization configurations and the corresponding 
off-shell action. 
The off-shell 5d BPS equations for the vector multiplet sector were solved 
in~\cite{Gomes:2013cca},~\cite{Gupta:2019xac}.
As shown in those papers, the resulting localization manifold is parameterized by one 
real parameter per off-shell vector multiplet. 
The resulting configurations reduce consistently to the corresponding 4d solutions. 
As explained in~\cite{Gupta:2019xac}, this is the complete localization manifold in the 
five-dimensional vector multiplet sector with an appropriate choice of reality conditions on the fields. 
For the Weyl-multiplet sector the complete localization manifold is not known. Here we 
lift the 4d KK multiplet solution to obtain an off-shell solution in the 5d Weyl multiplet as 
in~\cite{Gomes:2013cca}. 

The presentation of this section is as follows. 
In Section~\ref{actvec} we evaluate the action on the off-shell vector multiplet solutions around 
the full-BPS (on-shell) AdS$_2$ Weyl multiplet background. 
In Section~\ref{offshellWeyl} we then move on to discuss the off-shell fluctuations 
in the Weyl-muliplet sector obtained  by lifting the 4d KK multiplet. 
Upon including the Weyl-multiplet off-shell fluctuation, the five-dimensional vector multiplet solutions 
around~AdS$_2$ also gets deformed. 
We then evaluate the off-shell action, including higher-derivative terms, on this manifold.

\subsection{Off-shell vector multiplets with on-shell Weyl background \label{actvec}}

We begin by writing all the fields as their attractor values, 
denoted by $*$, plus a fluctuation term,
\be
\sigma \= \sigma_{*} + \Sigma\,, \quad W \= W_{*} + w\,, \quad F \= F_{*} + f\,, \quad Y_{ij} \= 0 + y_{ij}\,.
\ee
For a suitable reality condition the non-zero fluctuations from the attractor values 
of vector multiplet fields found in~\cite{Gupta:2019xac} are
\be \label{solfluc}
\Sigma \= \frac{C\cosh\alpha}{\cosh\eta}\,, \quad \quad 
w_{\rho} \= -\i \, \frac{C\sinh\alpha\cosh\alpha}{\cosh\eta}\,, \quad\quad y_{12}\, \equiv  
k_{3} \= \frac{C}{2\cosh^{2}\eta}\,. 
\ee
The reality condition,~$W_\rho$ is taken to be imaginary, is consistent with the 
4d/5d connection. 
As described in section~\ref{EleQEF}, we evaluate the various terms in the action. 
Using equations~\eqref{Eqn:5dLVVV} and~\eqref{Eqn:LTDVVV}, we get the second derivative action,
\be
S_{VVV} - S_{TD} \= 2\pi \, \mathcal{C}_{IJK}
\Bigl(\frac{\sigma^{I}_{*}\sigma^{J}_{*}\sigma^{K}_{*}}{\cosh\alpha}(\cosh\eta_{0} - 1) 
-3\sigma^{I}_{*}C^{J}C^{K}\cosh\alpha - C^{I}C^{J}C^{K}\cosh^2\alpha\Bigr)\,,
\ee
where we have introduced a large cut-off,~$\eta_0$.
The higher derivative Lagrangian~\eqref{Eqn:5dLVWW}, and the total derivative 
term~\eqref{Eqn:LTDVWW}, evaluated on the off-shell vector multiplets is
\be
S_{VWW}- S^{TD}_{VWW} \= -\frac{\pi }{2} c_{I}\sigma^{I}_{*}(\cosh\eta_{0} -1)\cosh\alpha\,,
\ee
Note that the terms proportional to the fluctuations,~$\Sigma^I$ cancel out and this solution 
is the same as in the semi-classical case.
In order to renormalise the bulk action, we add the boundary term~\eqref{Eqn:SbulkBnd}, 
evaluated on the near-horizon of the black hole
Putting the three terms together, we get
\be
S_{\text{Bulk}} \= -\frac{2\pi \CC(\varphi)}{1+(\varphi^{0}_{*})^{2}} 
+ \frac{\pi  \ic_{I}\varphi^{I}}{2(1+(\varphi^{0}_{*})^{2})}\,,
\ee
where $\varphi^0_{*}$ is defined by~\eqref{Eqn:Newvarstar} and we define
\be \label{Eqn:NewvarI}
\varphi^{I} \= \frac{\sigma^{I}_{*}}{\cosh\alpha}+C^{I}\,.
\ee
The renormalised Wilson line is the sum of $S_{W.L.} + S^{2}_{\text{BND}}$, 
given by~\eqref{Eqn:SWL} and~\eqref{Eqn:WLBnd}, respectively. It evaluates 
on the BH solution including vector multiplet fluctuations to give 
\be
S_{W.L.}^{\text{Ren}} \= \pi q_{I}\varphi^{I} + \pi J \varphi^{0}_{*}\,.
\ee
Adding all of the terms together we get
\be
S(q_{I}, J; \varphi^{I}, \varphi^{I}_{*}) \= \pi q_{I}\varphi^{I} 
+ \pi J \varphi^{0}_{*}-2\pi \CC(\varphi)\bigl(\frac{1}{1+(\varphi^{0}_{*})^{2}}) 
+ \frac{\pi}{2} \ic_{I}\varphi^{I}\bigl(\frac{1}{1+(\varphi^{0}_{*})^{2}}\bigr)\,.
\label{Eqn:stotalV}
\ee
We now wish to evaluate the action over the fluctuations of the Weyl multiplet. 

\subsection{Off-shell Weyl multiplet \label{offshellWeyl}}

We now lift the localization manifold from~4d to~5d, including the fluctuation of the Kaluza-Klein vector multiplet. 
The 4d and 5d constants are related by
\be
C^{\widehat{I}} \= 2C_{4}^{\widehat{I}}\,.
\ee
Now we wish to lift this to fluctuations of the Weyl multiplet in 5d. This problem was discussed 
in~\cite{Gomes:2013cca} where it was found that the lift of the KK multiplet solves the 5d Weyl 
multiplet localization equations. We lift the 4d solutions to write the 5d solutions in the Weyl multiplet. 
The 4d/5d connection was discussed in~\eqref{Eqn:4d5dviel} where the Kaluza-Klein direction is called~$\rho$. 
Upon lifting the~$\wh I =0$ solution in~\eqref{4dlocsols} to~5d, we obtain
\bea
e^{a}_{\mu} & \= & e_{4d}^{a}{}_{\mu}\,, \\
\phi & \;\equiv \; & e^{\rho}{}_{5} \= \frac{\sqrt{1+ (\Sigma^{0})^{2} + 2i\Sigma^{0}\sinh \alpha}}{\cosh\alpha}\,, \\
B_{\mu} &\; \equiv \;& \phi e^{5}{}_{\mu} \= \{-\tanh\alpha(\cosh\eta -1),0,0,\cos\psi\}\,, \\
V_{\rho}{}^{1}_{1} &\= & - V_{\rho}{}^{2}_{2} 
 \= -\frac{\Sigma^{0}\cosh \alpha}{\cosh\eta\sqrt{1+ (\Sigma^{0})^{2} + 2i\Sigma^{0}\sinh \alpha}}\,, \\    
V_{\mu}{}^{1}_{1} & \=  & -V_{\mu}{}^{2}_{2} \=  V_{\rho}{}^{1}_{1}B_{\mu}\,.
\eea
The auxiliary field $T$ has the following fluctuations,
\be
\begin{split}
T_{12}& \= \frac{- \i \, \cosh\alpha}{4\sqrt{1+ (\Sigma^{0})^{2} + 2i\Sigma^{0}\sinh \alpha}}\,,  \\
T_{34} & \= \frac{\i \, \sinh\alpha}{4\sqrt{1+ (\Sigma^{0})^{2} + 2i\Sigma^{0}\sinh \alpha}} 
+ \frac{\Sigma^{0}}{3\sqrt{1+ (\Sigma^{0})^{2} + 2i\Sigma^{0}\sinh \alpha}}\,, \\
T_{25} & \= \frac{\Sigma^0 \tanh\eta \cosh\alpha}{6(1+ (\Sigma^{0})^{2} + 2i\Sigma^{0}\sinh \alpha)}\,.
\end{split}
\ee
The auxiliary field~$D$ is 
\be
D \= -\frac{(\Sigma^{0})^{2} \bigl(\tanh^2\eta + (\Sigma^{0})^2+1\bigr)}{12 \bigl((\Sigma^{0})^2+1\bigr)^2}\,.
\ee

\vspace{0.4cm}

In the presence of the Weyl multiplet fluctuations, the localization solutions for the vector multiplet
also change. Using the 4d/5d connection, discussed in section~\ref{sec:4d5dfields}, we obtain 
\be
\begin{split}
& \sigma^{I}  \=  \frac{\sigma^{I}_{*} + \Sigma^{I}}{\sqrt{1+ (\Sigma^{0})^{2} + 2i\Sigma^{0}\sinh \alpha}}\,, \qquad 
\\
& Y^{I}{}^{12}  \=  \frac{\Sigma^{I} - (\Sigma^{0})^2\sigma^{I}_{*} 
- i \Sigma^{0}\sinh\alpha (\sigma^{I}_{*}-\Sigma^{I})}{2\cosh\alpha\cosh\eta(1 + (\Sigma^{0})^2 
+2\i \,\Sigma^0 \sinh\alpha)} \,,\\
& W^I_{\rho}  \= -\frac{(\sigma^I_{*} + \Sigma^{I})(\i\,\sinh\alpha + \Sigma^{0})}{1+ (\Sigma^{0})^{2} 
+2 i\Sigma^{0}\sinh \alpha}\,,
\qquad W^{I}_{\mu} \=  A^{I}_{\mu} + W^{I}_{\rho}B_{\mu}\,.
\end{split}
\ee

We now evaluate the second derivative action over the localization manifold including Weyl multiplet fluctuations. 
We also add the renormalised Wilson line, including all fluctuations.  
The second derivative action is
\be
S_{\text{Ren}}( \vec{q}, J;\vec{\varphi}, \varphi^{0}) \=  \pi q_{I}\varphi^{I} 
+ \pi J \varphi^{0}-2\pi \Bigl(\frac{\CC(\varphi)}{1+(\varphi^{0})^{2}}\Bigr)\,,
\label{Eqn:stotal2W}
\ee
where we express the action in terms of the variables, $\varphi^{I}$, defined 
in~\eqref{Eqn:NewvarI} and
\be \label{Eqn:Newvar0}
\varphi^0 \= \i \tanh\alpha + C^{0}\,.
\ee
The action~\eqref{Eqn:stotal2W} agrees with the answer found in~\cite{Gomes:2013cca} and,
as we shall see, reduces to the corresponding~4d action.

So far we have obtained the two partial results for the 
action~\eqref{Eqn:stotalV} and~\eqref{Eqn:stotal2W}. Equation~\eqref{Eqn:stotalV}
corresponds to the full five-dimensional action evaluated over the vector multiplet fluctuations with the 
Weyl multiplet fixed to its on-shell value~$\v^0_*$. 
Equation~\eqref{Eqn:stotal2W} corresponds to the second derivative action evaluated over all
the vector multiplet and Weyl multiplet fluctuations. 
The minimal formula that combines these two expressions is 
\be
S_{\text{Ren}}( \vec{q},J;\vec{\varphi}, \vp^0) \=  \pi q_{I}\varphi^{I} 
+ \pi J \varphi^{0}-2\pi \Bigl(\frac{\CC(\varphi)}{1+(\varphi^{0})^{2}}\Bigr) 
+ \frac{\pi}{2} \Bigl(\frac{\ic_{I}\varphi^{I}}{1+(\varphi^{0})^{2}}\Bigr)\,,
\label{Eqn:stotalW}
\ee 
where the variables~$\vp^I$ and~$\vp^0$ are defined by~\eqref{Eqn:NewvarI} 
and~\eqref{Eqn:Newvar0} respectively.
Although we have not derived the five-dimensional total derivative terms in 
the higher-derivative part of the off-shell action---and this remains as a
technical gap that needs to be filled---it is difficult to see how to deform our 
proposal~\eqref{Eqn:stotalW} without destroying the symmetries of the problem
and consistency with the four-dimensional reduction. 
The reason this derivation is not trivial is that upon reducing the higher-derivative 
five-dimensional action to four dimensions, we find non-F-terms in, which have not been
completely classified, as we already discussed in Section~\ref{sec:nonF}.

In order to see that the action~\eqref{Eqn:stotalW} reduces precisely to the~4d action~\eqref{Eqn:stotal4d},
we note that the prepotential is related to the symmetric tensor of 5d as follows
\be
F\bigl(\frac{\phi+i p^I}{2}\bigr) - \overline{F}\bigl(\frac{\phi+i p^I}{2}\bigr) 
\= -\i \, \Bigl(\frac{\CC(\varphi)}{1+(\varphi^{0})^{2}}\Bigr) 
- \frac{\i}{4} \Bigl(\frac{\ic_{I}\varphi^{I}}{1+(\varphi^{0})^{2}}\Bigr)\,,
\ee
where we have used~\eqref{Eqn:fulleqSP} and~\eqref{Eqn:Prephd}.
We also remind the reader that the charges in~4d and~5d are the same up to a factor of four. 
Therefore, we see that
\be
S_{\text{Ren}} \= 4 \, S^{(4d)}_{\text{Ren}}\,.
\ee

In the next section we propose an ansatz for the one-loop determinant and combine 
it with \eqref{Eqn:stotalW} to obtain a formula for the quantum entropy function.

\section{Quantum entropy of the spinning black hole \label{sec:quantent}}

Upon putting together the results of the previous two sections, we obtain 
the localization result for quantum entropy, 
\be \label{Eqn:Zqu}
\begin{split}
Z(J,q^I) & \= \int\prod_{I = 0}^{N_{V}+1} \,d\varphi^I\,
\exp \bigl(S_{\text{Ren}}(\varphi^0, \varphi^I; J, q^I) \bigr) \,Z_{\text{1-loop}}(\varphi^0, \varphi^I; J, q^I)\,, \\
S_{\text{Ren}}(\varphi^0, \varphi^I; J, q^I) & \=  \pi q_{I}\varphi^{I} + \pi J \varphi^{0}
-2\pi  \frac{\CC(\varphi)}{1+(\varphi^{0})^{2}}  + \frac{\pi}{2} \frac{\ic_{I}\varphi^{I}}{1+(\varphi^{0})^{2}} \,.
\end{split}
\ee
In~\S~\ref{sec:one-loop}, we constrain the one-loop determinant using on-shell results for 
logarithmic corrections to the entropy.
In~\S~\ref{sec:topMthy} we compare our result to the one obtained from the topological M-theory conjecture.

\subsection{Logarithmic corrections to entropy and ansatz for $Z_{\text{1-loop}}$ \label{sec:one-loop}}

In Section~\ref{sec:Sclass1}, we calculated the semi-classical entropy using a saddle point approximation,
\be
\eS^{BH}_{*} \= \frac{1}{4} A_H \=2\pi \sqrt{Q^3 - \tfrac{1}{4}J^2}\,,
\ee
where $A_{H}$ is the area of the horizon.
The saddle points were found to be at
\be
\begin{split}
\varphi^{0}_{*} \= -\frac{J}{2\sqrt{Q^3-\frac{1}{4}J^{2}}}\,,\\
\varphi^{I}_{*} \= \widehat{q}_{I} \, \sqrt{\frac{Q^3}{Q^3 - \frac{1}{4}J^{2}}}\,, 
\end{split} \label{Eqn:Saddlepoint}
\ee
where we recall that 
\be
q_{I} \= 6 \, C_{IJK} \, \widehat{q}^{J} \, \widehat{q}^{K}\,, \qquad 
Q^{3/2} \defeq 2\, \CC(\widehat{q}) \,.
\ee
The leading correction to the area law is due to a term proportional to the logarithm of the area. 
The paper~\cite{Sen:2012cj} calculated the leading logarithmic correction by summing various 
one-loop diagram contributions in supergravity to the quantum entropy formula~\eqref{qefdef}.  
The result was that if charges scale as
\be
q_I  \sim \Lambda\,, \quad \quad 
J  \sim \Lambda^{3/2}\,,\quad\quad \Lambda \rightarrow \infty \,,
\ee \label{Eqn:scalecharge}
so that $A_{H} \sim \Lambda^\frac{3}{2}$, then the entropy 
is given by
\be 
\label{Eqn:LEL}
\eS^{\text{qu}} \= \begin{cases}
\eS^{BH}_{*}-\frac{1}{4}(n_{V}-3)\log \Lambda + O(\Lambda^{-1}) \,, \;  & J \neq 0\,,\\
\eS^{BH}_{*}-\frac{1}{4}(n_{V}+3)\log \Lambda+ O(\Lambda^{-1})\,,  &J= 0\,.
\end{cases}
\ee

\vspace{0.2cm}

We can now compare the low energy result~\eqref{Eqn:LEL} with the formula~\eqref{Eqn:Zqu} 
derived from localization, and calculate the scaling behavior of~$Z_{\text{1-loop}}$.
Upon performing a saddle point approximation to~\eqref{Eqn:Zqu}, we find that the partition function is
\be 
Z(\q,J) \= e^{S_{\text{Ren}}(\v_{*})} \, Z_{\text{1-loop}}(\v_{*}) \, \sqrt{\frac{2\pi}{|S_{\text{Ren}}''(\v_{*})|}} 
+ O(\Lambda^{-1})\,,
\ee
where $S_{\text{Ren}}''(\v_{*})$ is the Hessian matrix derived from Equation~\eqref{Eqn:stotalstar}.
Expanding this to second order we obtain the following result for the quantum entropy,
\be \label{Squ2ndorder}
S^{\text{qu}}(\q,J) \= \eS^{\text{BH}}_{*} + \log Z_{\text{1-loop}}(\v_*) 
-\frac{1}{2}\log |S_{\text{Ren}}''(\v_{*})| + \dots\,.
\ee
For the scaling of charges, \eqref{Eqn:scalecharge}, the saddle point equations \eqref{Eqn:Saddlepoint} 
tell us that 
\be \label{phistarscaling}
\v_{*} \; \sim \; \Lambda^{1/2}\,, \qquad \v_{*}^0 \; \sim \; \Lambda^{0} \,.
\ee
The second derivatives of the renormalised action scale as
\be
\begin{split}
(S_{\text{Ren}})_{IJ}(\v_{*}) &\= \frac{-12 \, \pi \, \CC_{IJK} \, \v^{K}_{*}}{1+(\v^{0}_{*})^2} \; 
\sim \; \Lambda^{\frac{1}{2}}\,,\\
(S_{\text{Ren}})_{I0}(\v_{*}) &\= \frac{12 \, \pi \,  \CC_{IJK} \, 
\v^{J}_{*}\v^{K}_{*}\v^{0}_{*}}{(1+(\v^{0}_{*})^2)^2}  \; \sim \;  \Lambda\,,\\
(S_{\text{Ren}})_{00}(\v_{*}) &\= \frac{4 \,\pi \, \CC(\v)(1-3(\v^{0}_{*})^2)}{(1+(\v^{0}_{*})^2)^3}  
\; \sim \;  \Lambda^{\frac{3}{2}}\,.
\end{split}
\ee
Therefore the determinant of the Hessian scales as
\be
|S_{\text{Ren}}''(\v_{*})| \sim \Lambda^{\frac{n_{V}+3}{2}}\,.
\ee
Comparing Equations~\eqref{Squ2ndorder} and~\eqref{Eqn:LEL}, we find that as~$\Lambda \to \infty$,
\be \label{Eqn:1loopsc}
\log Z_{\text{1-loop}} \; \sim \; \begin{cases}
\frac{3}{2}\, \log \Lambda  \quad & J \neq 0\,,\\
0 & J= 0\,.\\
\end{cases}
\ee
Assuming that the one-loop determinant is smooth near~$\v^0=0$, the symmetries of the problem 
(in particular, the invariance under rotation of the vector index~$I$) and the 
result~\eqref{Eqn:1loopsc} together lead to the following expression,
\be \label{1loopansatz}
Z_{\text{1-loop}} (\v) \= \begin{cases}
\CC(\v) \, f(\v^0) \; & J \neq 0\,,\\
f(\v^0)  & J = 0\,,\\
\end{cases}
\ee
where $f(\v^0)$ is a function of~$\v^0$ which cannot be fixed by scaling arguments 
(recall from~\eqref{phistarscaling} that $\v^0$ does not scale with~$\Lambda$).

We thus reach our quantum gravitational entropy formula~\eqref{Eqn:Zqu}, \eqref{1loopansatz}.
Before moving on to discuss the relation of this formula with the topological M-theory 
conjecture, we note that in the context of extended supersymmetry we have exact formulas 
for the microstate degeneracy of BPS BHs~\cite{Maldacena:1999bp},~\cite{Dabholkar:2010rm, Dabholkar:2012nd}.
We leave a detailed comparison for future work, but discuss some features of such a comparison using
a particular three-charge model in Appendix~\ref{sec:STUmodel}.

\subsection{5d black holes and the topological M-theory conjecture \label{sec:topMthy}}

We first recall some details of the set up of~\cite{Dijkgraaf:2004te} describing M-theory compactified 
on a Calabi-Yau three-fold~$M$.  
We begin with the non-rotating case~$J=0$. The black hole is then specified 
by the electric charge vector~$Q \in H_2(M,\IZ)$ interpreted as an M2-brane wrapping a 2-cycle of~$M$,
with Poincar\'e dual~$ [\s] \in H^4(M,\IZ)$.
The 4-form~$\s$ obeys the stability criterion of~\cite{Hitchin:2001rw}, which implies that it can be written 
as 
\be \label{ksigrel}
\s \= \frac12 \, k \wedge k \,,
\ee
in terms of a symplectic 2-form~$k$. The attractor mechanism implies 
that the moduli fields are fixed by~$k$, and that the black hole entropy is given by the Hitchin 
functional for~$\sigma$,
\be
V_S(\s) \= \frac16 \int_M k\wedge k \wedge k   \,.
\ee
Writing
\be \label{kphirel}
k = \v^I \alpha_I\,,
\ee
where~$\alpha_I$ are an integral basis of~$H^2(M,\IZ)$, the relation~\eqref{ksigrel}
leads to the familiar equations for the attractor moduli~$\v^I_*$ (see Equation~\eqref{qiqihatrel})
\be
q_{I} \= 6 \, C_{IJK} \, \v_*^J \, \v_*^K\,, 
\ee
where the intersection numbers are given by 
\be
6\, C_{IJK} \=  \int_M \alpha_I \wedge  \alpha_J \wedge  \alpha_K \,.
\ee
The attractor entropy, given in Equations~\eqref{SBHQ32}, is 
\be
S_\text{BH} \=  4 \pi \, C_{IJK} \, \v_*^I \, \v_*^J \, \v_*^K\,, 
\ee
is simply the classical value of the Hitchin functional  
\be
S_\text{BH} \= 4 \pi \, V_S(\s_*) \,,
\ee
or, in other words, the symplectic volume of the Calabi-Yau manifold.

Within this set up, the proposal of~\cite{Dijkgraaf:2004te} is that the 5d BH entropy 
is given by a functional integral governed by the Hitchin functional~$V_S$ 
\be \label{ZqZsRel}
Z(\vec{q})\; \equiv \; \text{e}^{\mathcal{S}^{\text{qu}(\vec{q})}} \; \sim \;  
Z_S([\s]) \; \= \; \int d \s \exp \bigl( 4 \pi \, V_S(\v) - \,q_I \, \v^I\bigr) \,.
\ee 
As explained in~\cite{Dijkgraaf:2004te}, this proposal is the A-model version (or M-theory)
of the conjecture. The B-model (or Type IIB) version is formulated in terms of the 
holomorphic Hitchin functional and yields the (perhaps more familiar) OSV formula 
which relates the 4d black hole entropy to the topological string partition function.

We put a~$\sim$ symbol in Equation~\eqref{ZqZsRel} because the proposal of DGNV 
did not fix the details of higher-derivative terms, the 1-loop determinant and 
the measure (or the correct integration variable). 
We can now compare this result to the gravitational entropy 
formula~\eqref{Eqn:Zqu} obtained by the 5d supergravity approach taken in this paper.  
We identify the integration variable to be~$\varphi$ (related to~$\s$ 
by~\eqref{ksigrel},~\eqref{kphirel}), and constrain the one loop determinant by~\eqref{1loopfinal}.
With these identifications, the gravitational entropy~\eqref{Eqn:Zqu} has the same 
form as~\eqref{ZqZsRel} for~$J=0$ when~$f(\v)=1$. 

For the rotating case~$J \neq 0$, the proposal of DGNV uses the functional
\be
V_S(\s, \v_J) \=  \int  \sqrt{\s^3 - (\v_J)^2} \,,
\ee
where they introduce the 6-form field\footnote{In the paper~\cite{Dijkgraaf:2004te} the field itself is 
called~$J$, but we do not use this notation since we use~$J$ for the value of the spin itself.}~$\v_J$.  
Here our quantum gravitational formula~\eqref{ZqZsRel}-which is consistent with the macroscopic 
on-shell one-loop results~\cite{Sen:2012cj} differs from DGNV (except of course in its classical value). 
As mentioned in the introduction, there is no square-root in our formula~\eqref{Eqn:Zqu}.
This difference is similar to the Nambu-Goto versus the Polyakov form of the action of a string worldsheet.

Finally, we remark that our gravitational derivation of the quantum entropy includes the 
higher-derivative term proportional to~$c_I$, which is the second Chern class of the~$CY_3$-fold.
Perhaps this suggests how higher derivative corrections can also be included into the a priori 
definition of topological M-theory.

\section*{Acknowledgements}

We would like to thank Atish Dabholkar, Valentin Reys, Cumrun Vafa, and Bernard de Wit for 
useful discussions and comments.
This work~is supported by the ERC Consolidator Grant N.~681908, ``Quantum black 
holes: A macroscopic window into the microstructure of gravity'',  by the STFC grant ST/P000258/1,
and by the ISIRD grant 9-406/2019/IITRPR/5480.

\appendix 

\section{Auxiliary fields and Curvature tensors in 4d/5d}\label{App:rem4d5dfields}
The auxiliary~$T$ field reduces as follows, 
\be \label{Eqn:4d5dT} 
T_{ab}\= \frac{-\i}{24|X^0|}(T^{-}_{ab}\overline{X}^{0} - F^{0-}_{ab})+h.c.\,, \quad \quad
T_{a5} \= \frac{\i}{12}\frac{D_{a}X^0}{X^{0}} + h.c.\,.
\ee
The auxiliary field~$D$ is given by
\be \label{Eqn:4d5dD}
D \= \frac{1}{4}(\widehat{D} - \frac{1}{4}\phi^{-1}D^{\mu}D_{\mu}\phi - \frac{\widehat{R}\phi}{24} - \frac{3\phi^{-2}}{32}F^{0}_{ab} F^{0ab}
-\frac{3}{2} T^{ab}T_{ab} -3T^{a5}T_{a5} - \frac{\phi^2}{4}V_{i}{}^{j}V_{j}{}^{i})\,.
\ee
The remaining fields reduce as
\be \label{Eqn:4d5dV}
B_{\mu} \= A^0_{\mu}\,,\quad\quad
V_{ai}{}^{j} \= \widehat{V}_{ai}{}^{j}\,, \quad \quad
V_{5}^{ij} \= -\frac{1}{2}\widehat{Y}^{0ij}|X^0|^{-1}\,.
\ee 
Lastly we have $b_{a} = \widehat{b}_{a}\,, b_{5} = 0$.

We also have from \cite{Banerjee:2011ts} how the curvature tensors reduce in 4d. Note due to a difference in convention, we have additional minus signs. We repeat here the ones we need
\be
\begin{split} \label{Eqn:4d5dR}
R_{ab}{}^{cd} & \= \widehat{R}_{ab}{}^{cd} - \frac{1}{2}\phi^{2}(F^{0}_{a}{}^{[c}F^{0}_{b}{}^{d]}+F^{0}_{cd}F^{0}_{ab})\,, \\
R_{ab}{}^{c5} & \= -\frac{1}{2}\phi^{-1}\eD^{c}F^{0}_{ab} +\phi^{-2}(\eD^{c}\phi F^{0}_{ab}-F^{0c}{}_{[a}\eD_{b]}\phi) \,,\\
R_{a5}{}^{b5} & \= \phi \eD_{a}(\phi^{-2}\eD^{b}\phi) + \frac{1}{4}F^{0}_{ac}F^{0}\,,\\
R_{ab} &\= \widehat{R}_{ab}-\frac{1}{2}\phi^{-2}F^{0}_{ac}F^{0}_{b}{}^{c} + \phi \eD_{a}(\phi^{-2}\eD_{b}\phi) \,, \\
R_{a5} &\= -\frac{1}{2}\phi^-1 \eD^{b}F^{0}_{ba} +\frac{3}{2}\phi^{-2}F^{0}_{ba}\eD^{a}\phi   \,,\\
R_{55} & \= \phi \eD_{a}(\phi^-2 \eD^a \phi) + \frac{1}{4}\phi^{-2}(F^{0})^2 \,,\\
R & \= \widehat{R} + 2\phi \eD_{a}(\phi^{-2}D^{a}\phi)-\frac{1}{4}\phi^{-2}(F^{0})^2\,.
\end{split}
\ee
We also note that
\be 
\begin{split}
\widetilde{R}_{\wedge} & \, \equiv \,\varepsilon^{5abcd}R_{ab}{}^{EF}R_{cdEF} \\
& \= \varepsilon^{abcd}\Bigl( \widehat{R}(M)_{ab}{}^{ef}\widehat{R}(M)_{cdef} +\phi^{-2}\widehat{R}_{ab}{}^{ef}(F^{0}_{ce}F^{0}_{df} + F^{0}_{cd}F^{0}_{ef})\\
& \qquad + \frac{1}{4}\phi^{-4}F^{0}_{ab}(2F^{0}_{ce}F^{0}_{df}F^{0ef} + F^{0}_{cd}(F^{0})^2) + \frac{1}{2}\phi^{-2} D_{e}F^{0}_{ab}D^{e}F^{0}_{cd}\\
& \qquad 2\phi^{-1}(D^{e}F^{0}_{ab})(F^{0}_{ce}D_{d}\phi^{-1} + F^{0}_{cd}D_{e}\phi^{-1})\\
& \qquad 2F^{0}_{ab}(F^{0}_{cd}(D_{e}\phi^{-1})^2 + 2 F^{0}_{ce}D^{e}(\phi^{-1}D_{d}\phi^{-1})\Bigr)\,.
\end{split}
\label{Eqn:RwedgeR}
\ee
and
\be \label{Eqn:4d5dR5wedge}
R_{5[a}{}^{EF}R_{cd]EF} \= \phi \mathcal{D}_{[a}\widetilde{R}_{cd]}\,,
\ee
where
\be
\begin{split}\label{Eqn:Rtilde}
\widetilde{R}_{cd}& \= \frac{\phi^{-2}}{2} R_{cd}{}^{ef}F^{0}_{ef}- \frac{\phi^{-4}}{8}((F^{0})^{2}F^{0}_{cd} + 2F^{0ef}F^{0}_{ce}F^{0}_{df})\\
& \quad \quad +2 \phi^{-1}F^{0}_{ce}\mathcal{D}_{d}\mathcal{D}^{e}\phi^{-1} - F^{0}_{cd}(D\phi^{-1})^2\,.
\end{split}
\ee

\section{Terms 1 and 2 of second derivative Lagrangian in 4d}\label{App:rem4d5dLag}
In order to reduce $L_{1}$, given by \eqref{Eqn:L1}, we first note using \eqref{Eqn:4d5dVec} the following
\be
\begin{split}
\frac{1}{2}D_M \sigma^I D^M \sigma^J &\= -\frac{|X^0|^2}{4}D_M t^{I}_{-}D^M t^{J}_{-}+\frac{|X^0|}{2} t^{I}_{-}D_M|X^0| D^M t^{J}_{-}\,, \\
\frac{1}{8}F^{I}_{AB}F^{JAB} & \= \frac{|X^0|^2}{4}D_M t^{I}_{+} D^M t^{J}_{+} +\frac{1}{8}\Bigl(\widehat{F}^{I}_{\mu\nu}\widehat{F}^{J\mu\nu} - t^{I}_{+}\widehat{F}^{J}_{\mu\nu}F^{0\mu\nu} +\frac{1}{4}t^{I}_{+}t^{J}_{+} F^{0}_{\mu\nu}F^{0\mu\nu} \Bigr)\,, \\
-\frac{1}{2}Y^{I}_{ij}Y^{Jij} & \= -\frac{1}{8}\Bigl(\widetilde{Y}^{I}_{ij}\widetilde{Y}^{Jij} - t^{I}_{+}\widetilde{Y}^{J}_{ij}Y^{0ij} +\frac{1}{4}t^{I}_{+}t^{J}_{+} Y^{0}_{ij}Y^{0ij} \Bigr)\,, \\
-\frac{3}{2}\sigma^{I} F^{J}_{AB} T^{AB} & \= \frac{1}{16}t^{I}_{-}\bigl[\bigl(\widehat{F}^{J}_{\mu\nu} -\frac{1}{2}t^{J}_{+}F^{0}_{\mu\nu}\bigr)\bigl(T^{-\mu\nu} \overline{X}^{0} - F^{-0\m\n} \bigr) -h.c. \bigr] \\
& \quad \quad + \frac{1}{4}t^{I}_{-} D_{M} t^{J}_{+}(\overline{X}^{0}D^{M} X^{0} -h.c. )\,.  
\end{split}
\ee
Putting this together, we find that $L_{1}$ reduces to
\be
\begin{split}
L_{1} & \= -3\i e|X^0|^2 \CC_{IJK}t^I_{-}\eD^{\mu}t^J\, \eD_{\mu}\overline{t}^K +\frac{3}{4}\i e\CC(t_{-})(D_{\mu}|X^0|)^2 \\
& \quad -\frac{3}{2}\i e \CC_{IJK}t^{I}_{-}t^{J}_{-}(\overline{X}^0 \eD^{\mu}\overline{t}^K\eD_{\mu}X^0 - h.c.) \\
&\quad -\frac{3}{8}\i e\CC_{IJK}t^{I}_{-}(F^J_{ab}F^{Kab} - t^{J}_{+}F^K_{ab}F^{0ab}+\frac{1}{4}t^{J}_{+}t^{K}_{+}(F^{0})^2))\\
& \quad +\frac{3}{8}\i e\CC_{IJK}t^{I}_{-}(Y^J_{ij}Y^{Kij} - t^{J}_{+}Y^K_{ij}Y^{0ij}+\frac{1}{4}t^{J}_{+}t^{K}_{+}|Y^{0}_{ij}|^2)\\
& \quad -\frac{3}{16}\i e \CC_{IJK}t^{I}_{-}t^{J}_{-}\left[(F^{K}_{ab}-\frac{1}{2}t^{K}_{+}F^{0}_{ab})(T^{-ab}\overline{X}^0-F^{0-ab}+h.c.)\right] 
\end{split}
\ee

We now calculate what $L_{2}$, \eqref{Eqn:L1}, reduces to in 4d. 
Using \eqref{Eqn:4d5dD}, \eqref{Eqn:4d5dVec} and the last equation in \eqref{Eqn:4d5dR},
we get
\be
\begin{split}
L_{2} & \= -\frac{\i e}{2|X^0|}\mathcal{C}(t_{-})\Bigl(\bigl(\frac{R_{4d}}{6} - D_{4d}\bigr)|X^0|^2 -\frac{1}{16}Y^{0}_{ij}Y_{0ij}  - |D_{M}X^0|^2 +\frac{3}{2}(D_{M}|X^0|)^2 \\
& \quad \quad + \frac{1}{32}F^{0}_{ab}F^{0}_{ab} + \frac{1}{32}\bigl(T^{-}_{ab} \overline{X}^0 -F^{0-}_{ab}  + h.c. \bigr)^2\Bigr)\,.
\end{split}
\ee
Given that the Lagrangians in 4d and 5d are related by~\eqref{Eqn:2derivstory} and~\eqref{Eqn:2derivstory}, the actions will be related by
\be \label{Eqn:Factorof4}
S^{5d} - S_{TD} \= 4 S^{4d}\,,
\ee
as the periodicity of the~$\rho$ coordinate is~$4\pi$.

\section{Evaluating the quantum entropy function for a three-charge model \label{sec:STUmodel}}

The microscopic degeneracies for 5d BHs in the context of extended supersymmetry in string theory
all essentially have the form of a Fourier coefficient of a modular (or mock modular) form. 
The Fourier coefficients are estimated, at all orders in perturbation theory in the large-charge limit,
by~$I$-Bessel functions (see e.g~\cite{Dijkgraaf:2000fq}, \cite{Dabholkar:2012nd, Bringmann:2012zr}). 

In this appendix we evaluate the quantum entropy for a particular model with three charges~$q_I$, $I=1,2,3$.
The model is related to the STU-model arising in string theory, which has played a central role in the  
comparison of micro and macro four-dimensional BH entropy~\cite{Dabholkar:2010rm, Dabholkar:2011ec}.
We show that the quantum entropy formula derived in Section~\ref{sec:quantent} applied to this 
model indeed produces a linear combinations of I-Bessel functions a expected. 

The model with three charges~$q_I$, $I=1,2,3$ is the following, 
\be
\ic_{I} \= 0\,, \quad \quad N_V \= 2\,, \quad \quad \CC(\v)  \= \v^1\v^2\v^3\,,
\ee
that is $\CC_{123}  \= 1/6$ and the rest of the components are equal by symmetry or set to vanish.
We take $I = 1$ to be the compensating vector multiplet. In addition, we fix the function~$f(\v^0)$
\be \label{fv0exp}
f(\v^0) \= \begin{cases}
\bigl(1+(\v^0)^2\bigr)^{-2}\,,\qquad J \,\neq\, 0\,, \\
\bigl(1+(\v^0)^2\bigr)^{-1}\,,\qquad J \= 0\,.
\end{cases}
\ee

\subsubsection*{Zero angular momentum $J = 0$}

We expand the quantum entropy function~\eqref{Eqn:Zqu} for the toy model ,
\be
Z \= \int d\v^1  d\v^0 d\v^2 d\v^3\, \frac{1}{1+(\v^0)^2}\,e^{\pi q_{I}\v^I   - \frac{2\pi\v^1}{(1+(\v^0)^2)}\v^2\v^3 }\,, 
\ee
where the ansatz for the one-loop determinant~\eqref{1loopansatz} is~$Z_{\text{1-loop}} = \bigl(1+(\v^0)^2\bigr)^{-1}$ when $J = 0$.
We evaluate the Gaussian integrals over $\v^2, \v^3$ to obtain
\be
Z \= -\i\int d\v^1  d\v^0 \, \frac{1}{\v^1} \,e^{\pi q_{1}\v^1+\frac{\pi q_2q_3}{2\v^1}(1+(\v^0)^2)}\,. 
\ee
The~$\v^0$ integral is also Gaussian (after a Wick-rotation). Upon integrating, we obtain
\be
Z \= \sqrt{\frac{2}{q_2 q_3}} \int d\v^1 \,\frac{1}{\sqrt{\v^1}}\,e^{\pi q_{1}\v^1 + \frac{\pi q_2 q_3}{2\v^1}}\,.
\ee
We can rewrite this partition function using the following integral representation of the~$I$-Bessel function
\be
\int dt \, t^{-(\rho + 1)} \, e^{at+\frac{b}{t}} \= 2\pi \i \, \bigl(\frac{a}{b}\bigr)^{\frac{\rho}{2}}\,I_{\rho}(\sqrt{4ab})\,.
\ee
We find that the partition function is
\be
Z \= \frac{2^{\frac{5}{4}}\pi \i}{(q_1q_2q_3)^{\frac{1}{4}}}I_{-\frac{1}{2}}(\sqrt{2\pi^2 q_1 q_2 q_3})\,.
\ee
Writing in terms of $Q^{\frac{3}{2}} = 2 C(\hat{q})$, we get $2Q^3 = q_1 q_2 q_3$, we obtain
\be
Z \= 2 \pi \i Q\,^{-\frac{3}{4}}\,I_{-\frac{1}{2}}\bigl(2\pi Q^\frac{3}{2}\bigr)\,.
\ee
To study the asymptotic behaviour of $\eS^{\text{qu}}$, we recall that the $Q \sim \Lambda$ 
for large~$\Lambda$. Therefore the partition function is
\be 
Z \= e^{S^{\text{qu}}}\, \sim\, e^{\Lambda^{\frac{3}{2}}}\Lambda^{-\frac{3}{2}} + \dots\,.
\ee
For the entropy, we asymptotically obtain
\be
\eS^{\text{qu}} \sim \Lambda^{\frac{3}{2}} - \frac{3}{2} \log\Lambda + \mathcal{O}(\Lambda^{-1})\,.
\ee
as expected by~\eqref{Eqn:LEL} for $n_V = N_{V} + 1 = 3$. 

\subsubsection*{Non-zero angular momentum for $J \neq 0$}
We now evaluate the quantum entropy function for the toy model when~$J \neq 0$. 
The one-loop determinant gives a pre-factor of~$\CC(\v)\bigl(1+(\v^0)^2\bigr)^{-1}$ to the integral,
\be
Z \= \int d\v^1  d\v^0 d\v^2 d\v^3\, \frac{\v^1 \v^2 \v^3}{(1+(\v^0)^2)^2}\,
e^{\pi q_{I}\v^I + \pi J \v^0  - \frac{2\pi\v^1}{(1+(\v^0)^2)}\v^2\v^3 }\,. 
\ee
The steps follow almost identically to the $J = 0$ case where the integrals can be 
simplified using the following Gaussian integral
\be
\int_{-\infty}^{\infty} \,dx\, x^2 e^{-a x^2} \= \bigl(\frac{1}{2a} +\frac{b^2}{4a^2}\bigr)\, 
\sqrt{\frac{\pi}{a}}e^{\frac{b^2}{4a}}\,.
\ee
The partition function is equal to
\be
Z \= a_{1} \, I_{-\frac{1}{2}}\Bigl(2\pi \sqrt{Q^3 - \frac{J^2}{4}}\Bigr) + a_{2} \, 
I_{\frac{1}{2}}\Bigl(2\pi \sqrt{Q^3 - \frac{J^2}{4}}\Bigr) + a_{3} \, I_{-\frac{3}{2}}\Bigl(2\pi \sqrt{Q^3 - \frac{J^2}{4}}\Bigr)\,, 
\ee
where the coefficients of the Bessel functions are
\be
a_1 \= \frac{i}{2} \Bigl(Q^3 - \frac{J^2}{4}\Bigr)^{-\frac{1}{4}}\,,\qquad 
a_2 \= \frac{i}{2}   \Bigl(Q^3 - \frac{J^2}{4}\Bigr)^{\frac{1}{4}}\,, \qquad 
a_3 \= \frac{i}{8}\, J^2\,  \Bigl(Q^3 - \frac{J^2}{4}\Bigr)^{-\frac{3}{4}}\,.
\ee
To see the asymptotic behaviour for~$\eS^{qu}$, we recall that~$J \sim \Lambda^\frac{3}{2}$ and so
\be
\begin{split}
a_{1}\,I_{-\frac{1}{2}}\Bigl(2\pi\sqrt{Q^3 - \frac{J^2}{4}}\Bigr)& \,\sim\, e^{\Lambda^{\frac{3}{2}}}\Lambda^{-\frac{3}{2}}\,, \\
a_2 \,I_{\frac{1}{2}}\Bigl(2\pi\sqrt{Q^3 - \frac{J^2}{4}}\Bigr)& \sim e^{\Lambda^{\frac{3}{2}}}\,, \\
a_3\,I_{-\frac{3}{2}}\Bigl(2\pi\sqrt{Q^3 - \frac{J^2}{4}}\Bigr)& \sim e^{\Lambda^{\frac{3}{2}}}\,.
\end{split}
\ee
The last two terms contribute to the leading order contribution to the entropy. For the entropy, this gives
\be
\eS^{\text{qu}} \sim \Lambda^{\frac{3}{2}} + \mathcal{O}(\Lambda^{-1})\,,
\ee
as expected by~\eqref{Eqn:LEL} for $n_V = 3$. 


\begin{thebibliography}{10}

\bibitem{Mandal:2010cj}
I.~Mandal and A.~Sen, {\it {Black Hole Microstate Counting and its Macroscopic
  Counterpart}},  {\em Nucl. Phys. Proc. Suppl.} {\bf 216} (2011) 147--168,
  [\href{https://arxiv.org/abs/1008.3801}{{\tt arXiv:1008.3801}}]. [Class.
  Quant. Grav.27,214003(2010)].


\bibitem{LopesCardoso:1998tkj}
G.~Lopes~Cardoso, B.~de~Wit, and T.~Mohaupt, {\it {Corrections to macroscopic
  supersymmetric black hole entropy}},  {\em Phys. Lett.} {\bf B451} (1999)
  309--316, [\href{https://arxiv.org/abs/hep-th/9812082}{{\tt
  hep-th/9812082}}].

\bibitem{Ooguri:2004zv}
H.~Ooguri, A.~Strominger, and C.~Vafa, {\it {Black hole attractors and the
  topological string}},  {\em Phys. Rev.} {\bf D70} (2004) 106007,
  [\href{https://arxiv.org/abs/hep-th/0405146}{{\tt hep-th/0405146}}].

\bibitem{Sen:2008vm}
A.~Sen, {\it {Quantum Entropy Function from AdS(2)/CFT(1) Correspondence}},
  {\em Int. J. Mod. Phys.} {\bf A24} (2009) 4225--4244,
  [\href{https://arxiv.org/abs/0809.3304}{{\tt arXiv:0809.3304}}].

\bibitem{Denef:2007vg}
F.~Denef and G.~W. Moore, {\it {Split states, entropy enigmas, holes and
  halos}},  {\em JHEP} {\bf 11} (2011) 129,
  [\href{https://arxiv.org/abs/hep-th/0702146}{{\tt hep-th/0702146}}].

\bibitem{Dabholkar:2010uh}
A.~Dabholkar, J.~Gomes, and S.~Murthy, {\it {Quantum black holes, localization
  and the topological string}},  {\em JHEP} {\bf 06} (2011) 019,
  [\href{https://arxiv.org/abs/1012.0265}{{\tt arXiv:1012.0265}}].

\bibitem{Dabholkar:2011ec}
A.~Dabholkar, J.~Gomes, and S.~Murthy, {\it {Localization \textbackslash{}\&
  Exact Holography}},  {\em JHEP} {\bf 04} (2013) 062,
  [\href{https://arxiv.org/abs/1111.1161}{{\tt arXiv:1111.1161}}].

\bibitem{Dabholkar:2014ema}
A.~Dabholkar, J.~Gomes, and S.~Murthy, {\it {Nonperturbative black hole entropy
  and Kloosterman sums}},  {\em JHEP} {\bf 03} (2015) 074,
  [\href{https://arxiv.org/abs/1404.0033}{{\tt arXiv:1404.0033}}].

\bibitem{Chowdhury:2019mnb}
A.~Chowdhury, A.~Kidambi, S.~Murthy, V.~Reys, and T.~Wrase, {\it {Dyonic black
  hole degeneracies in $\mathcal{N} = 4$ string theory from Dabholkar-Harvey
  degeneracies}},  {\em JHEP} {\bf 10} (2020) 184,
  [\href{https://arxiv.org/abs/1912.0656}{{\tt arXiv:1912.0656}}].

\bibitem{Dijkgraaf:2004te}
R.~Dijkgraaf, S.~Gukov, A.~Neitzke, and C.~Vafa, {\it {Topological M-theory as
  unification of form theories of gravity}},  {\em Adv. Theor. Math. Phys.}
  {\bf 9} (2005), no.~4 603--665,
  [\href{https://arxiv.org/abs/hep-th/0411073}{{\tt hep-th/0411073}}].

\bibitem{Hitchin:2000jd}
N.~J. Hitchin, {\it {The Geometry of Three-Forms in Six Dimensions}},  {\em J.
  Diff. Geom.} {\bf 55} (2000), no.~3 547--576,
  [\href{https://arxiv.org/abs/math/0010054}{{\tt math/0010054}}].

\bibitem{Hitchin:2001rw}
N.~J. Hitchin, {\it {Stable forms and special metrics}},
  \href{https://arxiv.org/abs/math/0107101}{{\tt math/0107101}}.

\bibitem{Ferrara:1996dd}
S.~Ferrara and R.~Kallosh, {\it {Supersymmetry and attractors}},  {\em Phys.
  Rev. D} {\bf 54} (1996) 1514--1524,
  [\href{https://arxiv.org/abs/hep-th/9602136}{{\tt hep-th/9602136}}].

\bibitem{Ferrara:1996um}
S.~Ferrara and R.~Kallosh, {\it {Universality of supersymmetric attractors}},
  {\em Phys. Rev. D} {\bf 54} (1996) 1525--1534,
  [\href{https://arxiv.org/abs/hep-th/9603090}{{\tt hep-th/9603090}}].

\bibitem{Breckenridge:1996is}
J.~C. Breckenridge, R.~C. Myers, A.~W. Peet, and C.~Vafa, {\it {D-branes and
  spinning black holes}},  {\em Phys. Lett.} {\bf B391} (1997) 93--98,
  [\href{https://arxiv.org/abs/hep-th/9602065}{{\tt hep-th/9602065}}].

\bibitem{Gupta:2019xac}
R.~K. Gupta, S.~Murthy, and M.~Sahni, {\it {On the localization manifold of 5d
  supersymmetric spinning black holes}},  {\em JHEP} {\bf 10} (2019) 172,
  [\href{https://arxiv.org/abs/1904.0887}{{\tt arXiv:1904.0887}}].

\bibitem{Murthy:2013xpa}
S.~Murthy and V.~Reys, {\it {Quantum black hole entropy and the holomorphic
  prepotential of N=2 supergravity}},  {\em JHEP} {\bf 10} (2013) 099,
  [\href{https://arxiv.org/abs/1306.3796}{{\tt arXiv:1306.3796}}].

\bibitem{Murthy:2015yfa}
S.~Murthy and V.~Reys, {\it {Functional determinants, index theorems, and exact
  quantum black hole entropy}},  {\em JHEP} {\bf 12} (2015) 028,
  [\href{https://arxiv.org/abs/1504.0140}{{\tt arXiv:1504.0140}}].

\bibitem{Gupta:2015gga}
R.~K. Gupta, Y.~Ito, and I.~Jeon, {\it {Supersymmetric Localization for BPS
  Black Hole Entropy: 1-loop Partition Function from Vector Multiplets}},  {\em
  JHEP} {\bf 11} (2015) 197, [\href{https://arxiv.org/abs/1504.0170}{{\tt
  arXiv:1504.0170}}].

\bibitem{Jeon:2018kec}
I.~Jeon and S.~Murthy, {\it {Twisting and localization in supergravity:
  equivariant cohomology of BPS black holes}},
  \href{https://arxiv.org/abs/1806.0447}{{\tt arXiv:1806.0447}}.

\bibitem{Gomes:2013cca}
J.~a. Gomes, {\it {Quantum entropy and exact 4d/5d connection}},  {\em JHEP}
  {\bf 01} (2015) 109, [\href{https://arxiv.org/abs/1305.2849}{{\tt
  arXiv:1305.2849}}].

\bibitem{Banerjee:2011ts}
N.~Banerjee, B.~de~Wit, and S.~Katmadas, {\it {The Off-Shell 4D/5D
  Connection}},  {\em JHEP} {\bf 03} (2012) 061,
  [\href{https://arxiv.org/abs/1112.5371}{{\tt arXiv:1112.5371}}].

\bibitem{Hanaki:2006pj}
K.~Hanaki, K.~Ohashi, and Y.~Tachikawa, {\it {Supersymmetric Completion of an
  R**2 term in Five-dimensional Supergravity}},  {\em Prog. Theor. Phys.} {\bf
  117} (2007) 533, [\href{https://arxiv.org/abs/hep-th/0611329}{{\tt
  hep-th/0611329}}].

\bibitem{Sen:2011ba}
A.~Sen, {\it {Logarithmic Corrections to N=2 Black Hole Entropy: An Infrared
  Window into the Microstates}},  {\em Gen. Rel. Grav.} {\bf 44} (2012), no.~5
  1207--1266, [\href{https://arxiv.org/abs/1108.3842}{{\tt
  arXiv:1108.3842}}].

\bibitem{Sen:2012cj}
A.~Sen, {\it {Logarithmic Corrections to Rotating Extremal Black Hole Entropy
  in Four and Five Dimensions}},  {\em Gen. Rel. Grav.} {\bf 44} (2012)
  1947--1991, [\href{https://arxiv.org/abs/1109.3706}{{\tt
  arXiv:1109.3706}}].

\bibitem{Castro:2007ci}
A.~Castro, J.~L. Davis, P.~Kraus, and F.~Larsen, {\it {Precision Entropy of
  Spinning Black Holes}},  {\em JHEP} {\bf 09} (2007) 003,
  [\href{https://arxiv.org/abs/0705.1847}{{\tt arXiv:0705.1847}}].

\bibitem{Castro:2008ne}
A.~Castro, J.~L. Davis, P.~Kraus, and F.~Larsen, {\it {String Theory Effects on
  Five-Dimensional Black Hole Physics}},  {\em Int. J. Mod. Phys. A} {\bf 23}
  (2008) 613--691, [\href{https://arxiv.org/abs/0801.1863}{{\tt
  arXiv:0801.1863}}].

\bibitem{deWit:2009de}
B.~de~Wit and S.~Katmadas, {\it {Near-Horizon Analysis of D=5 BPS Black Holes
  and Rings}},  {\em JHEP} {\bf 02} (2010) 056,
  [\href{https://arxiv.org/abs/0910.4907}{{\tt arXiv:0910.4907}}].

\bibitem{Castro:2008ys}
A.~Castro and S.~Murthy, {\it {Corrections to the statistical entropy of five
  dimensional black holes}},  {\em JHEP} {\bf 06} (2009) 024,
  [\href{https://arxiv.org/abs/0807.0237}{{\tt arXiv:0807.0237}}].

\bibitem{Dabholkar:2010rm}
A.~Dabholkar, J.~Gomes, S.~Murthy, and A.~Sen, {\it {Supersymmetric Index from
  Black Hole Entropy}},  {\em JHEP} {\bf 04} (2011) 034,
  [\href{https://arxiv.org/abs/1009.3226}{{\tt arXiv:1009.3226}}].

\bibitem{Dabholkar:2012nd}
A.~Dabholkar, S.~Murthy, and D.~Zagier, {\it {Quantum Black Holes, Wall
  Crossing, and Mock Modular Forms}},
  \href{https://arxiv.org/abs/1208.4074}{{\tt arXiv:1208.4074}}.

\bibitem{Gopakumar:1998ii}
R.~Gopakumar and C.~Vafa, {\it {M theory and topological strings. 1.}},
  \href{https://arxiv.org/abs/hep-th/9809187}{{\tt hep-th/9809187}}.

\bibitem{Gopakumar:1998jq}
R.~Gopakumar and C.~Vafa, {\it {M theory and topological strings. 2.}},
  \href{https://arxiv.org/abs/hep-th/9812127}{{\tt hep-th/9812127}}.

\bibitem{Cabo-Bizet:2018ehj}
A.~Cabo-Bizet, D.~Cassani, D.~Martelli, and S.~Murthy, {\it {Microscopic origin
  of the Bekenstein-Hawking entropy of supersymmetric AdS$_{5}$ black holes}},
  {\em JHEP} {\bf 10} (2019) 062,
  [\href{https://arxiv.org/abs/1810.1144}{{\tt arXiv:1810.1144}}].

\bibitem{Cassani:2019mms}
D.~Cassani and L.~Papini, {\it {The BPS limit of rotating AdS black hole
  thermodynamics}},  {\em JHEP} {\bf 09} (2019) 079,
  [\href{https://arxiv.org/abs/1906.1014}{{\tt arXiv:1906.1014}}].

\bibitem{Bobev:2019zmz}
N.~Bobev and P.~M. Crichigno, {\it {Universal spinning black holes and theories
  of class $ \mathcal{R} $}},  {\em JHEP} {\bf 12} (2019) 054,
  [\href{https://arxiv.org/abs/1909.0587}{{\tt arXiv:1909.0587}}].

\bibitem{BenettiGenolini:2019jdz}
P.~Benetti~Genolini, J.~M. Perez Ipi\~na, and J.~Sparks, {\it {Localization of
  the action in AdS/CFT}},  {\em JHEP} {\bf 10} (2019) 252,
  [\href{https://arxiv.org/abs/1906.1124}{{\tt arXiv:1906.1124}}].

\bibitem{Kantor:2019lfo}
G.~K\'antor, C.~Papageorgakis, and P.~Richmond, {\it {AdS$_{7}$ black-hole
  entropy and 5D $ \mathcal{N} $ = 2 Yang-Mills}},  {\em JHEP} {\bf 01} (2020)
  017, [\href{https://arxiv.org/abs/1907.0292}{{\tt arXiv:1907.0292}}].

\bibitem{David:2020ems}
M.~David, J.~Nian, and L.~A. Pando~Zayas, {\it {Gravitational Cardy Limit and
  AdS Black Hole Entropy}},  {\em JHEP} {\bf 11} (2020) 041,
  [\href{https://arxiv.org/abs/2005.1025}{{\tt arXiv:2005.1025}}].

\bibitem{Bobev:2020egg}
N.~Bobev, A.~M. Charles, K.~Hristov, and V.~Reys, {\it {The Unreasonable
  Effectiveness of Higher-Derivative Supergravity in AdS$_4$ Holography}},
  {\em Phys. Rev. Lett.} {\bf 125} (2020), no.~13 131601,
  [\href{https://arxiv.org/abs/2006.0939}{{\tt arXiv:2006.0939}}].

\bibitem{Bobev:2020zov}
N.~Bobev, A.~M. Charles, D.~Gang, K.~Hristov, and V.~Reys, {\it
  {Higher-Derivative Supergravity, Wrapped M5-branes, and Theories of Class
  $\mathcal{R}$}},  \href{https://arxiv.org/abs/2011.0597}{{\tt
  arXiv:2011.0597}}.

\bibitem{Dabholkar:2014wpa}
A.~Dabholkar, N.~Drukker, and J.~Gomes, {\it {Localization in supergravity and
  quantum $AdS_4/CFT_3$ holography}},  {\em JHEP} {\bf 10} (2014) 090,
  [\href{https://arxiv.org/abs/1406.0505}{{\tt arXiv:1406.0505}}].

\bibitem{Nian:2017hac}
J.~Nian and X.~Zhang, {\it {Entanglement Entropy of ABJM Theory and Entropy of
  Topological Black Hole}},  {\em JHEP} {\bf 07} (2017) 096,
  [\href{https://arxiv.org/abs/1705.0189}{{\tt arXiv:1705.0189}}].

\bibitem{Hristov:2018lod}
K.~Hristov, I.~Lodato, and V.~Reys, {\it {On the quantum entropy function in 4d
  gauged supergravity}},  {\em JHEP} {\bf 07} (2018) 072,
  [\href{https://arxiv.org/abs/1803.0592}{{\tt arXiv:1803.0592}}].

\bibitem{Hristov:2019xku}
K.~Hristov, I.~Lodato, and V.~Reys, {\it {One-loop determinants for black holes
  in 4d gauged supergravity}},  {\em JHEP} {\bf 11} (2019) 105,
  [\href{https://arxiv.org/abs/1908.0569}{{\tt arXiv:1908.0569}}].

\bibitem{deWit:2010za}
B.~de~Wit, S.~Katmadas, and M.~van Zalk, {\it {New supersymmetric
  higher-derivative couplings: Full N=2 superspace does not count!}},  {\em
  JHEP} {\bf 01} (2011) 007, [\href{https://arxiv.org/abs/1010.2150}{{\tt
  arXiv:1010.2150}}].

\bibitem{Butter:2013lta}
D.~Butter, B.~de~Wit, S.~M. Kuzenko, and I.~Lodato, {\it {New higher-derivative
  invariants in N=2 supergravity and the Gauss-Bonnet term}},  {\em JHEP} {\bf
  12} (2013) 062, [\href{https://arxiv.org/abs/1307.6546}{{\tt
  arXiv:1307.6546}}].

\bibitem{Ozkan:2013nwa}
M.~Ozkan and Y.~Pang, {\it {All off-shell $R^{2}$ invariants in five
  dimensional $\mathcal{N} =$ 2 supergravity}},  {\em JHEP} {\bf 08} (2013)
  042, [\href{https://arxiv.org/abs/1306.1540}{{\tt arXiv:1306.1540}}].

\bibitem{Baggio:2014hua}
M.~Baggio, N.~Halmagyi, D.~R. Mayerson, D.~Robbins, and B.~Wecht, {\it {Higher
  Derivative Corrections and Central Charges from Wrapped M5-branes}},  {\em
  JHEP} {\bf 12} (2014) 042, [\href{https://arxiv.org/abs/1408.2538}{{\tt
  arXiv:1408.2538}}].

\bibitem{Katz:1999xq}
S.~H. Katz, A.~Klemm, and C.~Vafa, {\it {M theory, topological strings and
  spinning black holes}},  {\em Adv. Theor. Math. Phys.} {\bf 3} (1999)
  1445--1537, [\href{https://arxiv.org/abs/hep-th/9910181}{{\tt
  hep-th/9910181}}].

\bibitem{Gaiotto:2005gf}
D.~Gaiotto, A.~Strominger, and X.~Yin, {\it {New connections between 4-D and
  5-D black holes}},  {\em JHEP} {\bf 02} (2006) 024,
  [\href{https://arxiv.org/abs/hep-th/0503217}{{\tt hep-th/0503217}}].

\bibitem{Skenderis:2002wp}
K.~Skenderis, {\it {Lecture notes on holographic renormalization}},  {\em
  Class. Quant. Grav.} {\bf 19} (2002) 5849--5876,
  [\href{https://arxiv.org/abs/hep-th/0209067}{{\tt hep-th/0209067}}].

\bibitem{Maldacena:1998im}
J.~M. Maldacena, {\it {Wilson loops in large N field theories}},  {\em Phys.
  Rev. Lett.} {\bf 80} (1998) 4859--4862,
  [\href{https://arxiv.org/abs/hep-th/9803002}{{\tt hep-th/9803002}}].

\bibitem{Gupta:2012cy}
R.~K. Gupta and S.~Murthy, {\it {All solutions of the localization equations
  for N=2 quantum black hole entropy}},  {\em JHEP} {\bf 02} (2013) 141,
  [\href{https://arxiv.org/abs/1208.6221}{{\tt arXiv:1208.6221}}].

\bibitem{Sen:2008}
A.~Sen, {\it {Black Hole Entropy Function, Attractors and Precision Counting of
  Microstates}},  {\em Gen. Rel. Grav.} {\bf 40} (2008) 2249--2431,
  [\href{https://arxiv.org/abs/0708.1270}{{\tt arXiv:0708.1270}}].

\bibitem{Marolf:2000cb}
D.~Marolf, {\it {Chern-Simons terms and the three notions of charge}},  in {\em
  {International Conference on Quantization, Gauge Theory, and Strings:
  Conference Dedicated to the Memory of Professor Efim Fradkin}}, pp.~312--320,
  6, 2000.
\newblock \href{https://arxiv.org/abs/hep-th/0006117}{{\tt hep-th/0006117}}.

\bibitem{Banerjee:2009uk}
N.~Banerjee, I.~Mandal, and A.~Sen, {\it {Black Hole Hair Removal}},  {\em
  JHEP} {\bf 07} (2009) 091, [\href{https://arxiv.org/abs/0901.0359}{{\tt
  arXiv:0901.0359}}].

\bibitem{Maldacena:1999bp}
J.~M. Maldacena, G.~W. Moore, and A.~Strominger, {\it {Counting BPS black holes
  in toroidal Type II string theory}},
  \href{https://arxiv.org/abs/hep-th/9903163}{{\tt hep-th/9903163}}.

\bibitem{Dijkgraaf:2000fq}
R.~Dijkgraaf, J.~M. Maldacena, G.~W. Moore, and E.~P. Verlinde, {\it {A Black
  hole Farey tail}},  \href{https://arxiv.org/abs/hep-th/0005003}{{\tt
  hep-th/0005003}}.

\bibitem{Bringmann:2012zr}
K.~Bringmann and S.~Murthy, {\it {On the positivity of black hole degeneracies
  in string theory}},  {\em Commun. Num. Theor Phys.} {\bf 07} (2013) 15--56,
  [\href{https://arxiv.org/abs/1208.3476}{{\tt arXiv:1208.3476}}].

\end{thebibliography}
\bibliographystyle{JHEP}

\providecommand{\href}[2]{#2}\begingroup\raggedright\endgroup

\end{document}